\documentclass[journal=jctcce,manuscript=article]{achemso}
\setkeys{acs}{articletitle=true}
\usepackage{graphicx}
\usepackage{amsmath}
\usepackage{amsfonts}
\usepackage{amssymb}
\usepackage{amsthm}
\usepackage{braket}
\usepackage{multirow}
\usepackage{achemso}
\usepackage{bm}%lettere greche in grassetto, e.g. $\bm{\rho}$
\usepackage{booktabs}
\usepackage{comment}
\usepackage{mathtools}% Loads amsmath
\usepackage{upgreek}

\usepackage[center]{subfigure}

\usepackage{adjustbox}
\usepackage{geometry}
\usepackage{multirow}
\usepackage[utf8]{inputenc}
\newlength\tindent
\newcommand*{\diff}{\mathop{}\!\mathrm{d}}             % per fare i differenziali
\newcommand{\tr}[1]{\text{Tr}{#1}} % abb. per scrivere sigma-cose in modo decente
\newcommand{\bh}[2]{\tr{\mathbf{h}^{#1}\mathbf{D}^{#2}}}
\newcommand{\bJ}[2]{\tr{\mathbf{D}^{#1}\mathbf{J}(\mathbf{D}^{#2})}}
\newcommand{\bK}[2]{\tr{\mathbf{D}^{#1}\mathbf{K}(\mathbf{D}^{#2})}}
\newcommand{\bex}[2]{\int \rho^{#1}(\mathbf{r})\varepsilon_x(\rho^{#2}(\mathbf{r}))\diff \mathbf{r}}
\newcommand{\bec}[2]{\int \rho^{#1}(\mathbf{r})\varepsilon_c(\rho^{#2}(\mathbf{r}))\diff \mathbf{r}}
\newcommand{\bexc}[2]{\int \rho^{#1}(\mathbf{r})\varepsilon_{xc}(\rho^{#2}(\mathbf{r}))\diff \mathbf{r}}

\newcommand{\bvx}[1]{\int v_{x}(\rho^{#1}(\mathbf{r}))\chi_\mu(\mathbf{r})\chi_\nu(\mathbf{r})\diff \mathbf{r}}
\newcommand{\bvc}[1]{\int v_{c}(\rho^{#1}(\mathbf{r}))\chi_\mu(\mathbf{r})\chi_\nu(\mathbf{r})\diff \mathbf{r}}
\graphicspath{{img/}} % Directory in which figures are stored

\author{Gioia Marrazzini$^{\top}$}
\affiliation{Scuola Normale Superiore,
             Piazza dei Cavalieri 7, 56126 Pisa, Italy.}  
\author{Tommaso Giovannini$^{\top}$}
\affiliation{Department of Chemistry, Norwegian University of Science and Technology, 7491 Trondheim, Norway}
\email{tommaso.giovannini@ntnu.no}
\author{Marco Scavino}
\affiliation{Scuola Normale Superiore,
             Piazza dei Cavalieri 7, 56126 Pisa, Italy.}  
\author{Franco Egidi}
\affiliation{Scuola Normale Superiore,
             Piazza dei Cavalieri 7, 56126 Pisa, Italy.}
\author{Chiara Cappelli}
\affiliation{Scuola Normale Superiore,
             Piazza dei Cavalieri 7, 56126 Pisa, Italy.}
\author{Henrik Koch}
\affiliation{Scuola Normale Superiore,
             Piazza dei Cavalieri 7, 56126 Pisa, Italy.}
\email{henrik.koch@sns.it}

\title[]
  {Multilevel Density Functional Theory}

\begin{document}

\begin{center}
$^{\top}$ G.M. and T.G. contributed equally to this work.
\end{center}

\begin{abstract}
We introduce a novel density-based multilevel approach in density functional theory. In this multilevel density functional theory (MLDFT), the system is partitioned in an active and an inactive fragment, and all interactions are retained between the two parts. In MLDFT, the Kohn-Sham equations are solved in the MO basis for the active part only, while keeping the inactive density frozen. This results in a reduction of computational cost. We outline the theory and implementation, and discuss applications to aqueous solutions of methyloxirane and glycidol.
\end{abstract}

\newpage

\section{Introduction}

The study of the energetics and physico-chemical properties of large molecular systems is one of the most challenging problems in quantum chemistry.\cite{dykstra2011theory} Many processes of chemical interest take place in solution,\cite{reichardt1992solvatochromism,buncel1990solvatochromism,
reichardt1994solvatochromic,cannelli2017understanding,
carlotti2018evaluation}  biological matrices\cite{cupellini2020charge,bondanza2020multiple} or at the interfaces between different materials.\cite{lomize2006positioning,furse2008dynamics,zhao2004high} %CHIARA and most of the spectroscopic measurements are performed in solution.
The large size of such systems poses theoretical and computational challenges because high-level correlated electronic structure methods are usually unfeasible due to their high computational cost and unfavorable scaling.\cite{myhre2016multilevel,hoyvik2017correlated} A good compromise between accuracy and computational cost is provided by density functional theory (DFT),\cite{parr1980density,burke2012perspective} which accounts for electron correlation in an approximate way. Due to the proven reliability of the results that can be obtained at the DFT level, it has become the most widely used approach for describing the electronic structure of large systems. 

Density functional theory permits the investigation of much larger systems than offered by highly correlated methods. However, it cannot be routinely applied to systems constituted by more than 500 atoms, unless implementations through graphical procession units (GPUs) are exploited.\cite{sisto2017atomistic} The practical limit of 500 atoms makes applications of DFT to biological matrices, interfaces and solutions particularly cumbersome and, in some cases, even impossible. For these reasons, several approximations have been developed in the past. 
Different approaches may be more or less suitable depending on the specificities of the system/environment couple. In the special case of systems in solution, particular success has been enjoyed by methods belonging to the family of the so-called focused models,\cite{tomasi2005,Mennucci12_386,tomasi1994molecular,
cappelli2016integrated} which are extremely useful when dealing with the property of a moiety or a chromophore embedded in an external environment.%, for instance a biological matrix or a solution.

In focused models, the target molecule is described at a higher level of theory with respect to the environment, which acts as a perturbation on the target system. Among the different focused models that have been developed in the past, the large majority belongs to the family of quantum mechanics (QM)/classical approaches, in which the target is treated at the QM level. The environment is instead described classically, by means of either continuous descriptions, such as the polarizable continuum model (PCM),\cite{tomasi2005,Mennucci12_386} or by retaining its atomistic nature in the so-called QM/molecular mechanics (QM/MM) approaches.\cite{warshel1972calculation,warshel1976theoretical,senn2009qm,lin2007qm} In all these methods, however, the interaction between the two parts of the whole system is usually described by classical electrostatics,\cite{curutchet2009electronic,olsen2011molecular,cappelli2016integrated,
giovannini2019fqfmu} and very rarely by including the interactions of quantum nature, such as Pauli repulsion and dispersion.\cite{giovannini2017disrep,giovannini2019eprdisrep,giovannini2019quantum}
Also, QM/classical methods allow for the treatment of very large systems, however their accuracy crucially depends on the quality of the parametrization of the classical fragments. In order to avoid such a variability, quantum embedding methods can be exploited.\cite{gordon2013accurate,gordon2007effective,sun2016quantum,
knizia2013density,chulhai2018projection,chulhai2017improved,
wen2019absolutely,ding2017embedded,knizia2013density,
goodpaster2012density,goodpaster2014accurate,manby2012simple,
goodpaster2010exact,zhang2020multi,ramos2015performance,
pavanello2011modelling} In these approaches, the whole system is treated by resorting to a QM description. The reduction in the computational cost is achieved by partitioning the system in at least one active and one inactive part. The former is accurately described, whereas the density of the latter is kept frozen, or described at a lower level of accuracy. Different approaches have been proposed in the past, ranging from projection-based methods, such as DFT-in-DFT or HF-in-DFT,\cite{bennie2017pushing,goodpaster2012density,
lee2019projection} or frozen density embedding (FDE).\cite{neugebauer2005merits,wesolowski2015frozen,fux2010accurate,
jacob2008flexible,jacob2006calculation,jacob2006comparison} %To mention some of such approaches, This for instance defined the so-called frozen density embedding, however a similar theoretical framework is common to most of the quantum embedding methods.

In this paper, we are proposing a novel quantum embedding approach defined in a DFT framework. We denote this method multilevel DFT (MLDFT), due to its similarity with multilevel Hartree-Fock (MLHF),\cite{saether2017density,hoyvik2020convergence} that we have recently developed.\cite{saether2017density} The MLDFT conceptually differs from the aforementioned quantum embedding methods because it is defined in the MO basis of the active fragment only. This feature automatically allows for a saving in the computational cost, because the inactive MOs are not involved in the self consistent field (SCF) procedure. In this paper, we derive MLDFT, and we apply the method to ground state energies of aqueous solutes. The results are compared, in all cases, to full DFT, in order to assess the quality of the multilevel partition. 

The manuscript is organized as follows. In the next section, DFT theory is formulated in the MO basis and MLDFT equations are presented and discussed with particular focus on the computational savings that can be expected. Then, after a brief section reporting on the computational details of the method, MLDFT is applied to selected aqueous systems, with emphasis on comparison with full DFT results. Conclusions and perspectives of the present work end the manuscript.

\section{Theory}

%\subsection{DFT in the Molecular Orbitals basis}
Our starting point is the DFT expression for the electronic energy of the system:
\begin{align}
E & = \bh{}{} + \dfrac{1}{2}\bJ{}{} - \dfrac{1}{2}c_x \bK{}{} + (1+c_x)E_x + Ec \nonumber \\
  & = \bh{}{} + \dfrac{1}{2}\bJ{}{} - \dfrac{1}{2}c_x \bK{}{} \nonumber \\
  & + (1-c_x)\bex{}{} + \bec{}{} \ .
\label{eq:dft_ao}
\end{align}
Here $\mathbf{D}$ is the density matrix, $\mathbf{h}$ is the one-electron operator, whereas $\textbf{J}$ and $\textbf{K}$ are coulomb and exchange matrices, respectively. The $E_{x}$ and $E_c$ terms are DFT exchange and correlation energies; $\rho(\mathbf{r})$ is the DFT density and $\varepsilon_{x}, \varepsilon_c$ are the exchange and correlation energy densities, respectively. 
The coefficient $c_{x}$ defines whether pure DFT ($c_x = 0$), or hybrid DFT functionals ($c_x \neq 0$) are exploited. 

The DFT density $\rho(\mathbf{r})$ is expressed in terms of the density matrix as:

\begin{equation}
\rho(\mathbf{r}) = \sum_{\mu\nu} D_{\mu\nu} \chi_\mu(\mathbf{r})\chi_\nu(\mathbf{r}) \ ,
\end{equation}

where, $\{\chi_{\mu}\}$ are the atomic orbitals (AO) basis functions. The energy defined in Eq. \ref{eq:dft_ao} is usually minimized in the AO basis. In order to reformulate the minimization in the MO basis, the same strategy developed for the Hartree-Fock case by Saether et al.\cite{saether2017density} can be used. This can be accomplished by parametrizing the density matrix $\mathbf{D}$ in terms of an antisymmetric rotation matrix, in which only the non-redundant occupied-virtual rotations are considered.\cite{saether2017density} 

\subsection*{Multilevel DFT}

The multilevel DFT (MLDFT) method belongs to the family of the so-called focused models. The part of the system which is under investigation (active) is described accurately, whereas the remaining (inactive) part remains frozen during the optimization of the active fragment.
The choice of the partitioning intimately depends on the specificities of the system, its chemical nature, and the properties one wishes to simulate.
Whatever the choice, within the MLDFT formalism the separation of the system into the two layers is based on the following decomposition of the density $\mathbf{D}$ and $\rho(\mathbf{r})$:

\begin{equation}
\textbf{D} = \textbf{D}^A + \textbf{D}^B \ , \qquad \qquad \rho(\mathbf{r}) = \rho^A(\mathbf{r}) + \rho^B(\mathbf{r}) \ ,
\label{eq:mldft-dens}
\end{equation}

where, $A$ and $B$ indicate the active and inactive fragments, respectively. As stated above, the active and inactive densities are usually defined on a physico-chemical basis. In case of a molecular system in solution, it is natural to define the solute as the active fragment, whereas the solvent molecules are treated as the inactive part. Notice, however, that the partitioning in Eq. \ref{eq:mldft-dens} is arbitrary and strongly depends on the method which is selected to mathematically decompose the total density matrix $\mathbf{D}$. In this work, a Cholesky decomposition of the total density is performed for the active occupied MOs, from which the active density $\mathbf{D}^A$ is calculated.\cite{aquilante2011cholesky,sanchez2010cholesky,koch2003reduced,saether2017density} The procedure ensures the all active and inactive orbitals are orthogonal. 

Now using Eq. \ref{eq:mldft-dens}, the total electronic energy in Eq. \ref{eq:dft_ao} can be written as:
%
%\begin{align}
%E & = \bh{}{A} + \bh{}{B} \nonumber \\
%  & + \dfrac{1}{2}\bJ{A}{A} + \dfrac{1}{2}\bJ{B}{B} + \dfrac{1}{2}\bJ{A}{B} + \dfrac{1}{2}\bJ{B}{A} \nonumber \\
%  & - c_x\left(\dfrac{1}{2}\bK{A}{A} + \dfrac{1}{2}\bK{B}{B} + \dfrac{1}{2}\bK{A}{B} + \dfrac{1}{2}\bK{B}{A}\right) \nonumber \\
%  & + (1-c_x)\bex{}{} + \bec{}{} \ .
%\end{align}
%
%
%
\begin{align}
E & = \bh{}{A} + \bh{}{B} \nonumber \\
  & + \dfrac{1}{2}\bJ{A}{A} + \dfrac{1}{2}\bJ{B}{B} + \bJ{A}{B} \nonumber \\
  & - c_x\left(\dfrac{1}{2}\bK{A}{A} + \dfrac{1}{2}\bK{B}{B} + \bK{A}{B} \right) \nonumber \\
  & + (1-c_x)\bex{}{} + \bec{}{} \ ,
\label{eq:MLDFT_iniziale}
\end{align}
where the symmetry of \textbf{J} and \textbf{K} matrices have been used.
Differently from MLHF,\cite{saether2017density} in MLDFT, the last term is not linear in the densities of the two subsystems. Therefore, we cannot directly separate it in contributions arising from $\rho^A$ and $\rho^B$. In order to get a physical understanding of eq. \ref{eq:MLDFT_iniziale}, we rewrite the last two terms by using this trivial identity for the exchange-correlation energy density ($\varepsilon_{xc} =  \varepsilon_x + \varepsilon_c$):

\begin{align}
 & \ \ \ \int [ \rho^A(\mathbf{r}) + \rho^B(\mathbf{r}) ] \varepsilon_{xc}(\rho^A(\mathbf{r}) + \rho^B(\mathbf{r}))  \diff \mathbf{r} \nonumber \\ 
 & =\int [ \rho^A(\mathbf{r}) + \rho^B(\mathbf{r}) ] \varepsilon_{xc}(\rho^A(\mathbf{r}) + \rho^B(\mathbf{r}))  \diff \mathbf{r} \nonumber \\
 & + \bexc{A}{A} - \bexc{A}{A} \nonumber \\
 & + \bexc{B}{B} - \bexc{B}{B} \nonumber \\
 & + \bexc{A}{B} - \bexc{A}{B} \nonumber \\
 & + \bexc{B}{A} - \bexc{B}{A} \ .
\label{eq:integrale_xc}
\end{align}

Substituting Eq. \ref{eq:integrale_xc} into Eq. \ref{eq:MLDFT_iniziale}, and reorganizing the different terms we obtain:

\begin{align}
E & = \bh{}{A} + \dfrac{1}{2}\bJ{A}{A} - \dfrac{1}{2}c_x\bK{A}{A} \nonumber \\
  & + (1-c_x)\bex{A}{A} + \bec{A}{A} \nonumber \\
  & + \bh{}{B} + \dfrac{1}{2}\bJ{B}{B} - \dfrac{1}{2}c_x\bK{B}{B} \nonumber \\
  & + (1-c_x)\bex{B}{B} + \bec{B}{B} \nonumber \\
  & + \bJ{A}{B} - c_x \bK{A}{B} \nonumber \\
  & + (1-c_x)\left(\bex{A}{B} + \bex{B}{A}\right) \nonumber \\
  & + \bec{A}{B} + \bec{B}{A} + E^{AB}_{non-add}\ ,
  \label{eq:ene_MLDFT} 
\end{align}

where, 

\begin{align}
E^{AB}_{nonadd} = (1-c_x)& \left( \bex{}{} - \bex{}{A} - \bex{}{B}\right) \nonumber \\
               & + \bec{}{} - \bec{}{A} - \bec{}{B}\ .
\end{align}

In Eq. \ref{eq:ene_MLDFT} the first four lines define the energy of the active and inactive fragments, whereas the last three lines define the active-inactive interaction. 
In MLDFT the density of the inactive part $\mathbf{D}^B$ (and $\rho^B(\mathbf{r})$) is frozen, and therefore it acts as an external field on the active fragment. Also, the energy terms containing the $B$ labels only are fixed during the SCF procedure.
%
% Fock
%
The total DFT Fock matrix is given by: 

\begin{align}
F_{\mu\nu} & = h_{\mu\nu} + J_{\mu\nu}(\textbf{D}) - c_x K_{\mu\nu}(\textbf{D}) \nonumber \\
 &+ (1-c_x) \bvx{} + \bvc{}\ ,
\label{eq:fock_tot}
\end{align}

where, $v_x(\rho(\mathbf{r}))$ and $v_c(\rho(\mathbf{r}))$ are the exchange and correlation potential densities, respectively. Using the partitioning in Eq. \ref{eq:mldft-dens}, Eq. \ref{eq:fock_tot} we get:
\begin{align}
F_{\mu\nu} & = h_{\mu\nu} + J_{\mu\nu}(\mathbf{D}^A) + J_{\mu\nu}(\mathbf{D}^B) - c_x \left( K_{\mu\nu}(\mathbf{D}^A) + K_{\mu\nu}(\mathbf{D}^B) \right) \nonumber \\
            & + (1 - c_x) \int v_{x} (\rho^A(\mathbf{r}) + \rho^B(\mathbf{r}))  \chi_\mu (\mathbf{r}) \chi_\nu(\mathbf{r}) \diff \mathbf{r} \nonumber \\
            & + \int v_{c} (\rho^A(\mathbf{r}) + \rho^B(\mathbf{r}))  \chi_\mu (\mathbf{r}) \chi_\nu(\mathbf{r}) \diff \mathbf{r} \ .
\label{eq:fock_AB}
\end{align}
We exploit the same identity of Eq. \ref{eq:integrale_xc} for the exchange-correlation potential density ($v_{xc} = v_x + v_c$). In this way, the last two terms in Eq. \ref{eq:fock_AB} become:

\begin{align}
& \ \ \int v_{xc} (\rho_A(\mathbf{r}) + \rho_B(\mathbf{r}))  \chi_\mu (\mathbf{r}) \chi_\nu(\mathbf{r}) \diff \mathbf{r} \nonumber \\
& = \int v_{xc} (\rho_A(\mathbf{r}) + \rho_B(\mathbf{r}))  \chi_\mu (\mathbf{r}) \chi_\nu(\mathbf{r}) \diff \mathbf{r} + \nonumber \\
& + \int v_{xc} (\rho_A(\mathbf{r}))  \chi_\mu (\mathbf{r}) \chi_\nu(\mathbf{r}) \diff \mathbf{r} - \int v_{xc} (\rho_A(\mathbf{r}))  \chi_\mu (\mathbf{r}) \chi_\nu(\mathbf{r}) \diff \mathbf{r} + \nonumber \\
& + \int v_{xc} (\rho_B(\mathbf{r}))  \chi_\mu (\mathbf{r}) \chi_\nu(\mathbf{r}) \diff \mathbf{r} - \int v_{xc} (\rho_B(\mathbf{r}))  \chi_\mu (\mathbf{r}) \chi_\nu(\mathbf{r}) \diff \mathbf{r}\ .
\end{align}

Reorganizing the terms in Eq. \ref{eq:fock_AB}, we can obtain the  working expression for the MLDFT Fock matrix:
\begin{align}
F_{\mu\nu} = h_{\mu\nu} & + \underbrace{J_{\mu\nu}(\textbf{D}^A) - c_x K_{\mu\nu}(\textbf{D}^A) + (1-c_x) \bvx{A} + \bvc{A}}_{2e_A}  \nonumber \\
           & + \underbrace{J_{\mu\nu}(\textbf{D}^B) - c_x K_{\mu\nu}(\textbf{D}^B) + (1-c_x) \bvx{B} + \bvc{B}}_{2e_B}  \nonumber \\
           & + \underbrace{(1-c_x) \int \chi_\mu(\mathbf{r})\chi_\nu(\mathbf{r}) [ v_{x}(\rho^A(\mathbf{r}) + \rho^B(\mathbf{r})) - v_{x}(\rho^A(\mathbf{r})) - v_{x}(\rho^B(\mathbf{r}))] \diff \mathbf{r}}_{2e^{non-add}_{AB}} \nonumber \\
           & + \underbrace{\int \chi_\mu(\mathbf{r})\chi_\nu(\mathbf{r}) [ v_{c}(\rho^A(\mathbf{r}) + \rho^B(\mathbf{r})) - v_{c}(\rho^A(\mathbf{r})) - v_{c}(\rho^B(\mathbf{r}))] \diff \mathbf{r}}_{2e^{non-add}_{AB}} \ ,
\label{eq:fock_MLDFT_final}
\end{align}
where, the two-electron contributions of A and B fragments and the interaction term AB are highlighted as $2e_{X}$, $\{X = A, B, AB\}$. 

There are two main advantages of using MLDFT compared to full DFT. Firstly, the HF exchange contribution is usually the most expensive term in most hybrid functionals. In MLDFT, only the active exchange term is to be computed at each SCF cycle, whereas the exchange integral of the inactive fragment is computed at the first SCF cycle only, as it is constant during the optimization. Second, the MLDFT SCF procedure can be performed in the MO basis of the active part only, thus intrinsically reducing the computational time as previously observed for the MLHF method.\cite{saether2017density}

\section{Computational Details}

The DFT and MLDFT are implemented in a development version of the electronic structure program e$^{\mathcal{T}}$ v.1.0.\cite{eT_arxiv}. In particular, the DFT grid is constructed using the widely employed Lebedev grid,\cite{lebedev1999quadrature} and the DFT functionals are implemented using the LibXC library.\cite{marques2012libxc}\\

A MLDFT calculation follows this computational protocol:

\begin{enumerate}
\item Construction of the initial density by means of superposition of molecular densities,\cite{neugebauer2005merits,he2010divide} followed by a diagonalization of the initial Fock matrix.
\item Partitioning of the new density into $A$ and $B$ densities, using Cholesky decomposition for the active occupied orbitals and projected atomic orbitals (PAOs) for active virtual orbitals.\cite{aquilante2011cholesky,sanchez2010cholesky,koch2003reduced}
\item Calculation of the constant energy terms and the one-electron contributions due to the inactive density $B$ entering in Eqs. \ref{eq:ene_MLDFT} and \ref{eq:fock_MLDFT_final}.
\item Minimization of the energy defined in Eq. \ref{eq:ene_MLDFT} in the MO basis of the active part $A$ only, until convergence is reached. 
\end{enumerate}

To show the robustness of MLDFT, three different functionals are used: LDA,\cite{kohn1965self} GGA (PBE\cite{perdew1997generalized}) and hybrid (B3LYP\cite{becke1993density}). These are combined with three different basis sets: 6-31G, 6-31G* and aug-cc-pVDZ.

\section{Numerical Applications}

In this section, the MLDFT is applied to some test cases to show the accuracy and the performance of the method. Solvation is one of the main physico-chemical phenomena in which such approaches can be exploited. We show the results of coupling MLDFT with two alternative, fully atomistic, strategies to model aqueous solutions. The first consists of a static modeling, which uses small clusters composed of the solute and a small number of surrounding water molecules. As an alternative, we apply MLDFT to snapshots extracted from a molecular dynamics (MD) simulation. In the latter framework, the dynamical aspects of the solvation phenomenon are retained, as are those arising from the combination of conformational changes in the solute and the surrounding solvent. In addition, long range interactions are taken into account. This latter modeling of the solvation phenomenon has been amply and successfully exploited by some of us within the framework of QM/MM approaches.\cite{giovannini2019quantum,giovannini2020csr,
giovannini2019fqfmulinear,cappelli2016integrated}

In the following sections the combination of MLDFT to the two aforementioned solvation approaches is tested, with application to two relatively small molecules, i.e. methyloxirane and glycidol in aqueous solution, which have been studied in the literature both theoretically and experimentally.\cite{giovannini2018effective,lipparini2013gauge,giovannini2019fqfmuir,
giovannini2017polarizable,giovannini2019fqfmu,losada2008solvation,
yang2009probing,merten2013anharmonicity,su2007hydration,
yu2009new,su2006chiral,su2006conformational} Such systems are chosen not only for their simplicity, but also because methyloxirane is a rigid molecule, whereas glycidol is not. Therefore, in the latter case, the results depend on: the selected QM level, and the approach used to solvation and conformational flexibility, which is instead discarded in the case of methyloxirane. In this way, we can dissect the various effects and highlight the quality of the MLDFT approach in details.

%Different solute+solvent clusters were recovered from the literature or constructed as previously reported. DOVE??????REFS!!!!!!!! For all the studied systems, MLDFT energies are compared to their full DFT counterparts.
%In this section, MLDFT energies of methyloxirane+water and glycidol+water clusters are reported and compared to full DFT results. Such clusters were selected because they have been previously used as sampling structures for the so-called cluster models for solvation. REFS!!!! As a final proof-of-concept calculation, we will show the performance of MLDFT in the reproduction of fullDFT solvation energies of different clusters of methyloxirane and glycidol and 50 water molecules extracted from Molecular Dynamics (MD) simulations.

\subsection{Cluster Models}

\subsubsection{Methyloxirane/water clusters}

\begin{figure}[htbp!]
\centering
\includegraphics[width=1\textwidth]{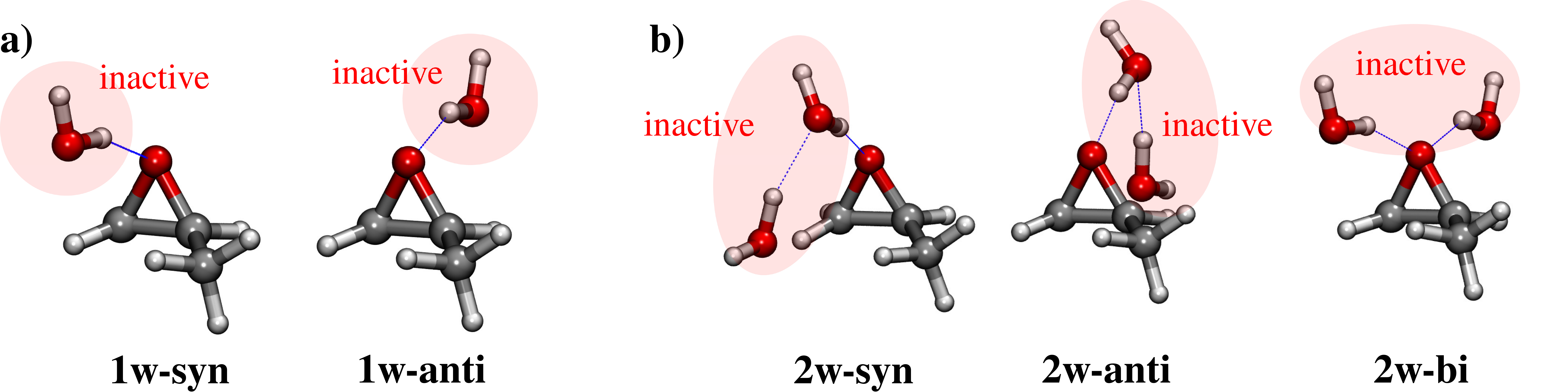}
\caption{Structure of conformers methyloxirane/water clusters. MLDFT partition is constructed so that Methyloxirane is the active part whereas the water molecules are the inactive fragments.}
\label{fig:moxy_struc}
\end{figure}

%\begin{figure}[htbp!]
%\centering
%\includegraphics[width=.6\textwidth]{po_2w}
%\caption{\textbf{2w-syn} (left), \textbf{2w-anti} (middle) and \textbf{2w-anti} (right) conformers of methyloxirane + 2 water cluster. MLDFT partition is constructed %so that Methyloxirane is the active part whereas the two water molecules are the inactive fragments.}
%\label{fig:moxy_2w}
%\end{figure}

%errore = 510^-4\%
%mindiff=pbe/6-31G (0.003137547362393)
%maxdiff=pbe/aug (0.354542852842261)
%CHIARA: MERGE Figs and 2-4 

The first studied solute is methyloxirane (MOXY), which is one of the smallest molecules that exhibits a chiral carbon. %For this reason, it has been amply studied both theoretically and experimentally. Also, its chiroptical response strongly depends on the solvent where MOXY is dissolved. In particular, its aqueous solution has been amply investigated because water solvent can even change the sign of the measured Optical Rotation (OR) of MOXY.\cite{egidi2012toward,lipparini2013optical} 
We have selected different clusters constituted by MOXY and one or two water molecules (see Figure \ref{fig:moxy_struc}), that have been previously studied by Xu and co-workers\cite{su2007hydration} to explain the unique characteristics of MOXY in aqueous solution.\cite{su2006conformational,su2007hydration} 

The two different conformers for the cluster composed of MOXY and one water molecule (MOXY+1w) are depicted in Fig. \ref{fig:moxy_struc}a. In the \textbf{1w-syn} structure water interacts with MOXY through hydrogen bonding on the same side of the methyl group, whereas the opposite occurs for the \textbf{1w-anti} structure. In both cases, MOXY is the active fragment, and water is the inactive moiety in MLDFT calculations.

\begin{figure}[htbp!]
\centering
\subfigure[][MOXY+1w clusters]
{\includegraphics[width=.47\textwidth]{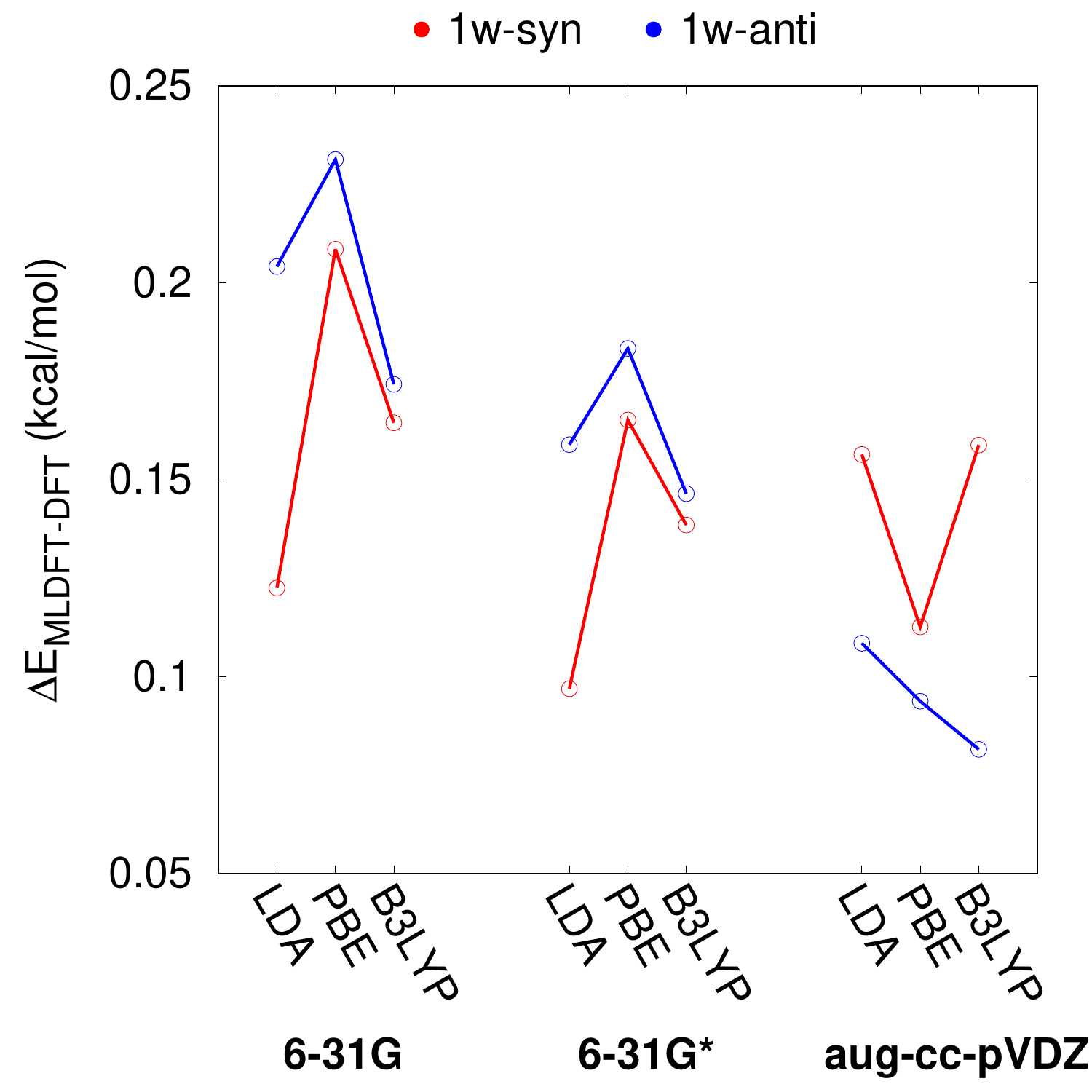}
\includegraphics[width=.47\textwidth]{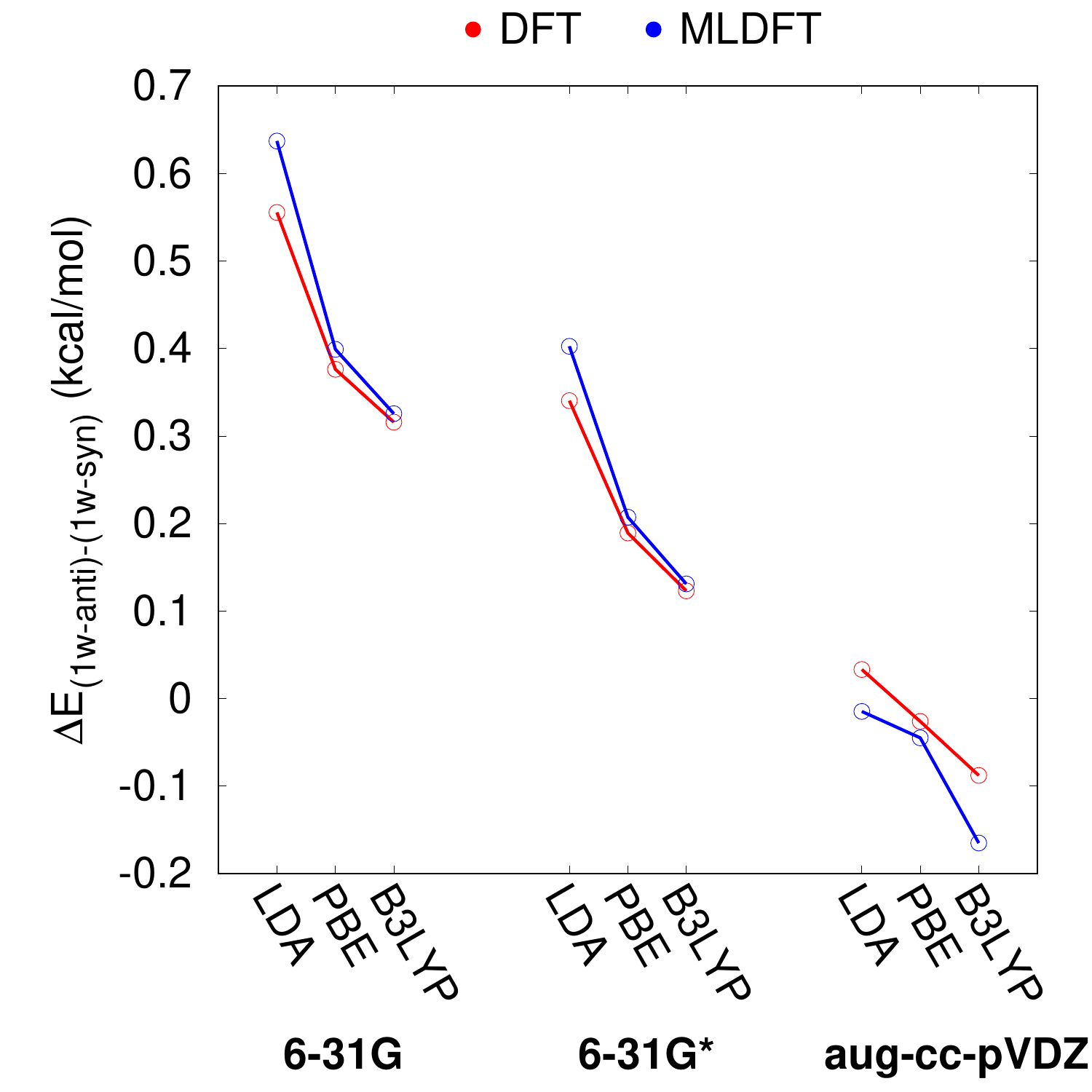}}
\subfigure[][MOXY+2w clusters]
{\includegraphics[width=.47\textwidth]{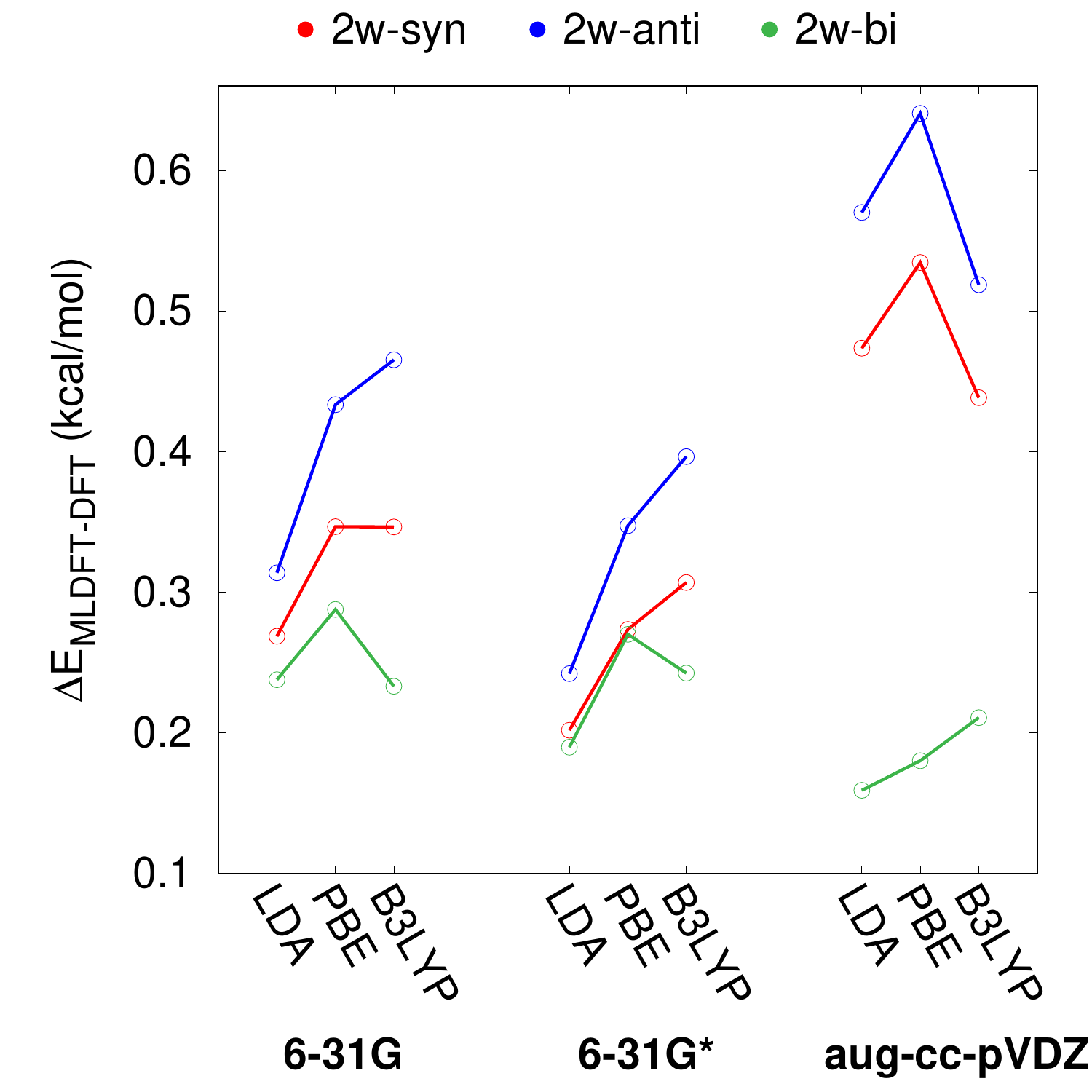}
\includegraphics[width=.47\textwidth]{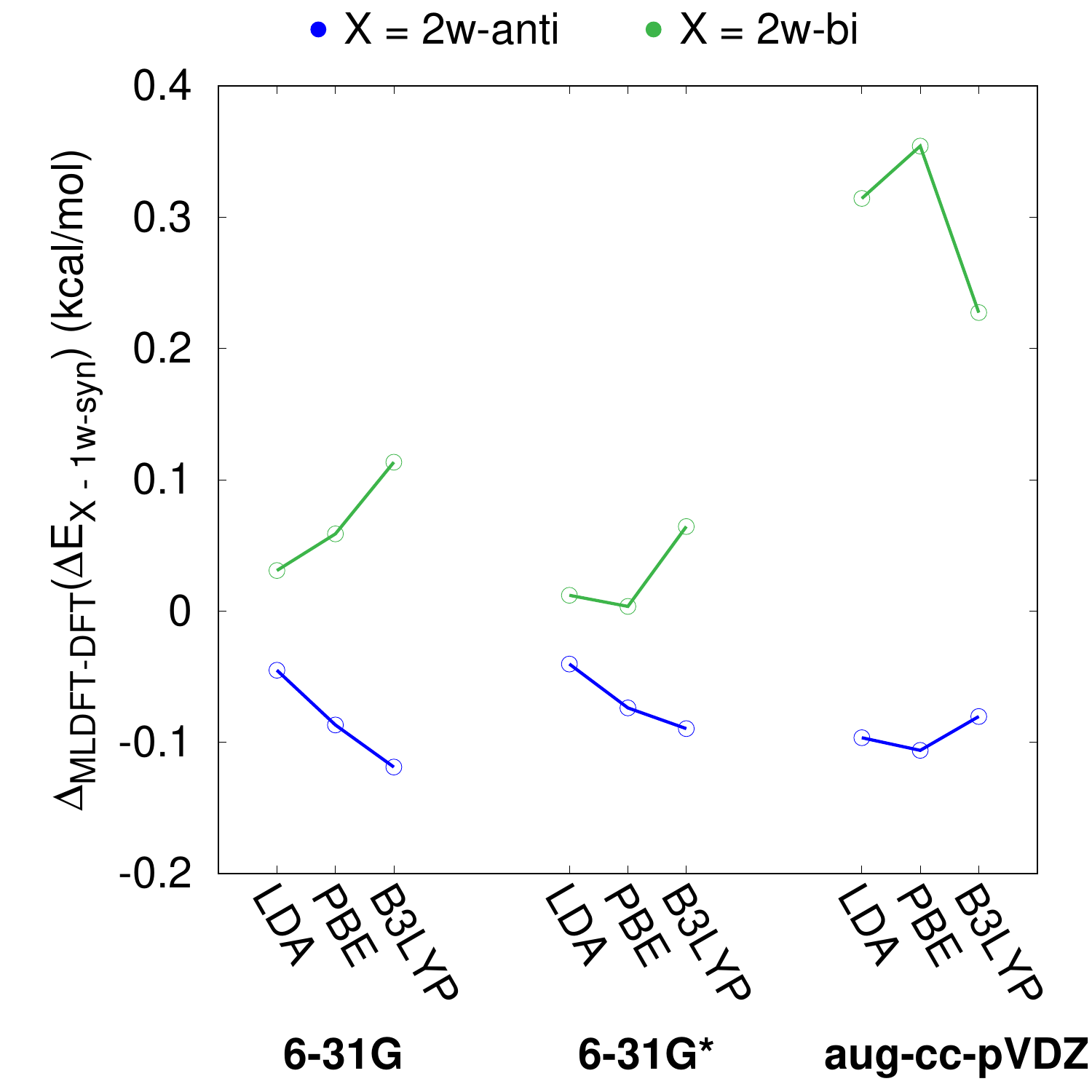}}
\caption{\textbf{(a)} (left) \textbf{1w-syn} and \textbf{1w-anti} total energy differences between MLDFT and DFT. (right) MLDFT and DFT energy difference between \textbf{1w-anti} and \textbf{1w-syn}. \textbf{(b)} (left) \textbf{2w-syn}, \textbf{2w-anti} and \textbf{2w-bi} total energy differences between MLDFT and DFT. (right) MLDFT-DFT difference on relative energies of \textbf{2w-anti} and \textbf{2w-bi} with respect to \textbf{2w-syn}.}
\label{fig:moxy_1w_2w_data}
\end{figure}

Ground state (GS) energy differences between DFT and MLDFT calculations are depicted in Fig. \ref{fig:moxy_1w_2w_data}, panel (a), left. 
%The LDA, PBE and B3LYP density functionals are used in combination with 6-31G, 6-31G* and aug-cc-pVDZ basis sets. 
Raw data are reported in Table S1 given in Supporting Information (SI). %CHIARA: CONTROLLA LE SI!!!!!!!!!!  
We see that the error between MLDFT and full DFT is below 1 mHartree ($<$ 0.628 kcal/mol), irrespective of the combination of functional/basis set employed. The error due to the MLDFT partitioning is well below the chemical accuracy (i.e. 1 kcal/mol).  %In particular, all computed errors between MLDFT and DFT are of the order of 10$^{-4}$\% on the total energies, thus clearly demonstrating the reliability of our method.

In the right panel of Fig. \ref{fig:moxy_1w_2w_data}a, DFT and MLDFT energy differences between \textbf{1w-anti} and \textbf{1w-syn} conformers are reported for all the considered combinations of functional/basis set. The raw data are reported in Tab. S1 in the SI. We see that DFT and MLDFT values almost coincide. In particular, LDA and PBE functionals predict \textbf{1w-syn} to be the most stable conformer, both at DFT and MLDFT level, independently of the selected basis set. Notice however that the energy difference between the two conformers decreases either as GGA functionals are employed or diffuse/polarization basis sets are used. The inclusion of HF exchange makes \textbf{1w-anti} the most stable conformer, if polarization/diffuse functions are considered. However, for all the considered combinations of functional/basis set, MLDFT and DFT values are almost perfectly in agreement, with the largest discrepancy being reported for B3LYP/aug-cc-pVDZ (0.08 kcal/mol). These findings clearly show that for this system MLDFT is able to catch small energy differences, which are again well below the chemical accuracy.

%\subsection{Methyloxirane + 2 water clusters}

We now turn to the clusters composed of MOXY and two water molecules (MOXY+2w, Figure \ref{fig:moxy_struc}b). Three main conformers are considered, according to Xu and co-workers:\cite{su2007hydration} \textbf{2w-syn}, \textbf{2w-anti} and \textbf{2w-bi}. The first two conformers differ from the position of water molecules, being both placed on the same side with respect to the methyl group in case of \textbf{2w-syn}, or on the opposite side for \textbf{2w-anti}. In \textbf{2w-bi} the two water molecules are instead placed on the opposite sides of the epoxyl oxygen atom. In all MLDFT calculations, MOXY is the active moiety, whereas the two water molecules are inactive.

In Fig. \ref{fig:moxy_1w_2w_data}b, left, GS energy differences between DFT and MLDFT for the three conformers are reported. The raw values associated with the data plotted in Fig. \ref{fig:moxy_1w_2w_data}b are given in Tab. S2 in the SI. The MLDFT and DFT results are, also in this case, in very good agreement, with an absolute error below 1 kcal/mol for all combinations of functional/basis sets. However, the absolute deviation between DFT and MLDFT energies is larger than for the previous case (see Fig. \ref{fig:moxy_1w_2w_data}a). In particular, MLDFT error is larger for \textbf{2w-syn} and \textbf{2w-anti} than for \textbf{2w-bi}, for which it is in line with what we have shown above for MOXY+1w clusters ($\sim$ 0.1 - 0.3 kcal/mol). The increase in the error may be justified by the fact that \textbf{2w-syn} and \textbf{2w-anti} feature one water molecule that is linked to another water molecule by means of intermolecular hydrogen bonding. The density of the inactive fragments (the two water molecules) is kept frozen, therefore the water molecule that is not directly bonded to the solute remains in its frozen electronic configuration, resulting in a larger error in the total energy. Such an hypothesis is confirmed by the fact that the error increases when the diffuse aug-cc-pVDZ basis set is used, and the same does not occur for \textbf{2w-bi}, where both water molecules are directly linked to methyloxirane through hydrogen bonding interactions.

The MLDFT-DFT deviations in energy differences between each conformer and \textbf{2w-syn} are shown in  Fig. \ref{fig:moxy_1w_2w_data}b, right. We note small discrepancies between MLDFT and full DFT, however also in this case they are below the chemical accuracy, with the maximum error reported by PBE/6-31G* ($\sim$ 0.35 kcal/mol). The error in the energy differences between the conformers is lower than for the total GS energies reported in Fig. \ref{fig:moxy_1w_2w_data}b, left.

\subsubsection{Glycidol/water clusters}

\begin{figure}[htbp!]
\centering
\includegraphics[width=.6\textwidth]{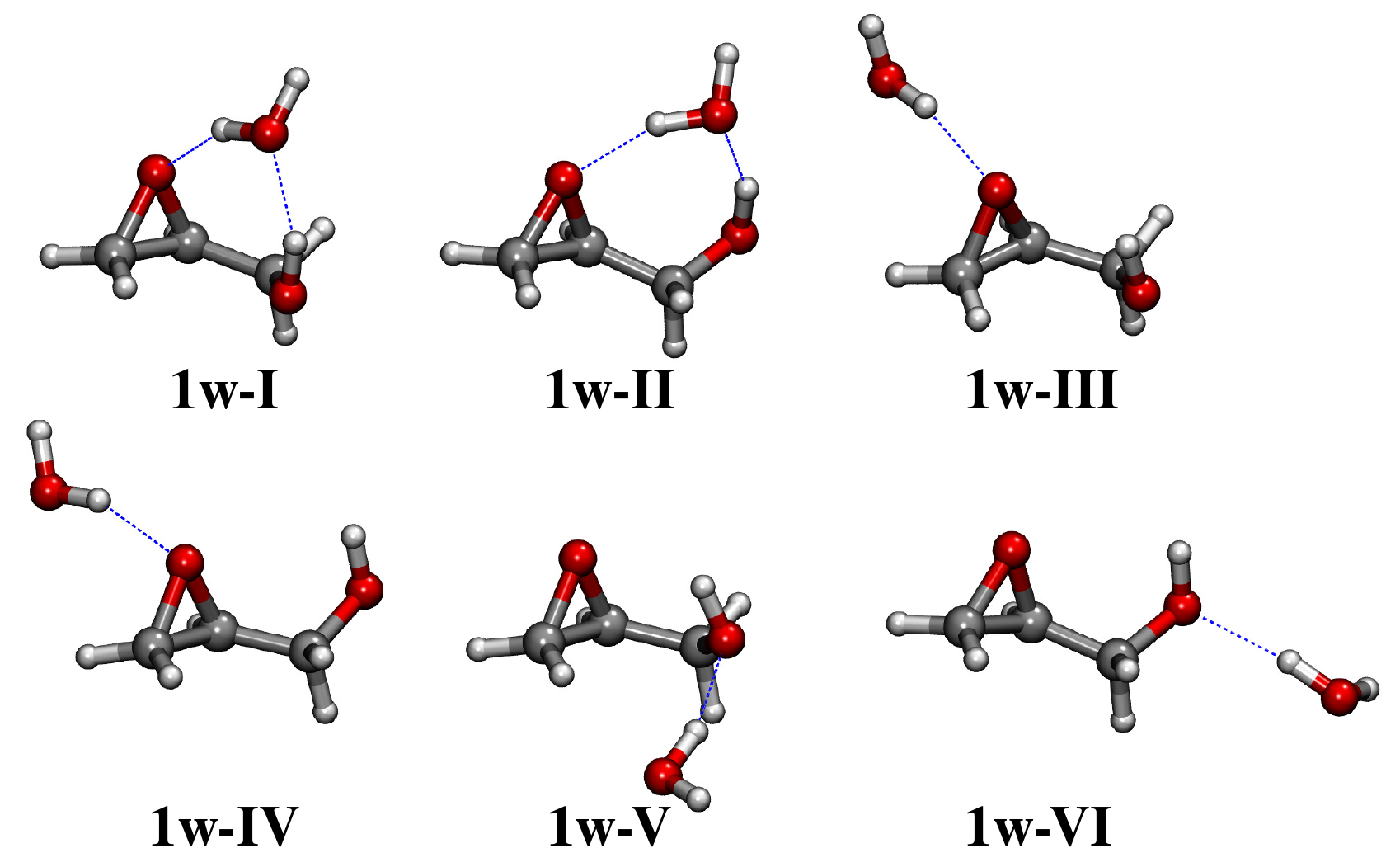}
\caption{Structures of the six conformers of glycidol + 1 water clusters. In MLDFT calculations, the glycidol moiety is active whereas the water molecule is inactive.}
\label{fig:gly_1w}
\end{figure}
\begin{figure}[htbp!]
\centering
\includegraphics[width=1\textwidth]{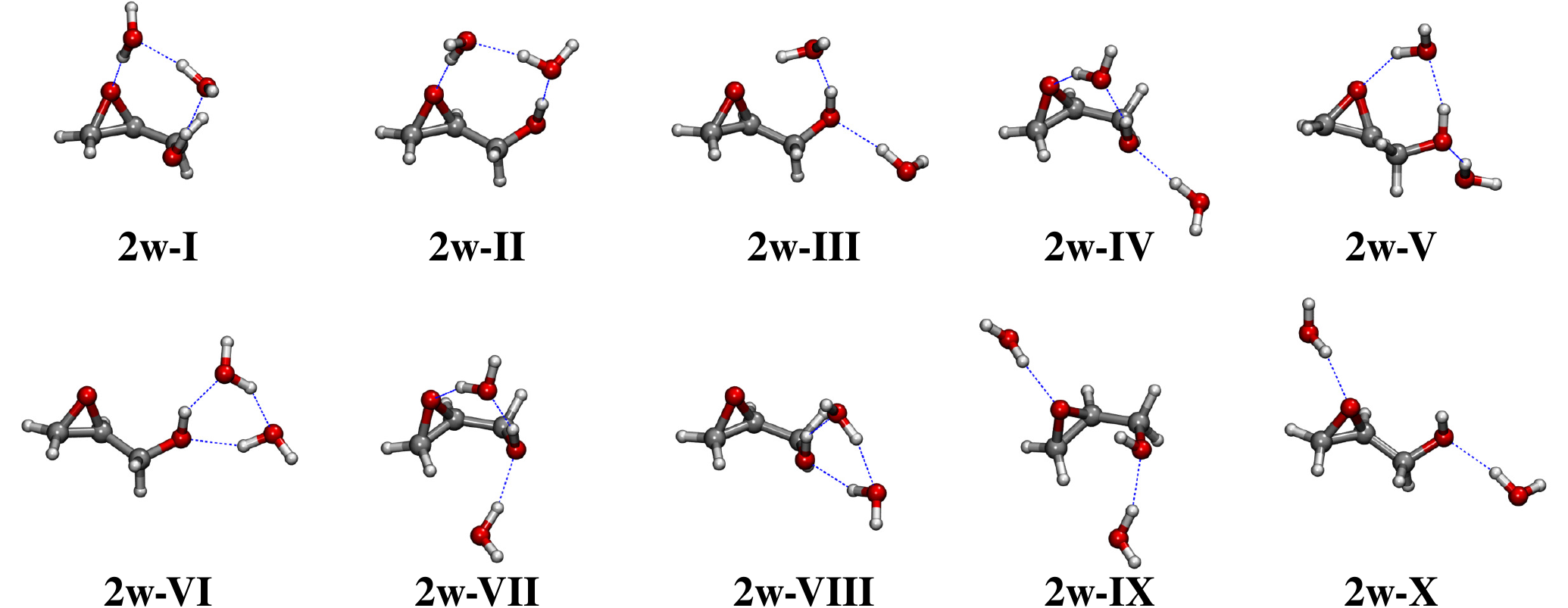}
\caption{Structures of the ten conformers of glycidol + 2 waters clusters. In MLDFT calculations,  glycidol is active whereas water molecules are inactive.}
\label{fig:gly_2w}
\end{figure}
\begin{figure}[htbp!]
\centering
\includegraphics[width=.7\textwidth]{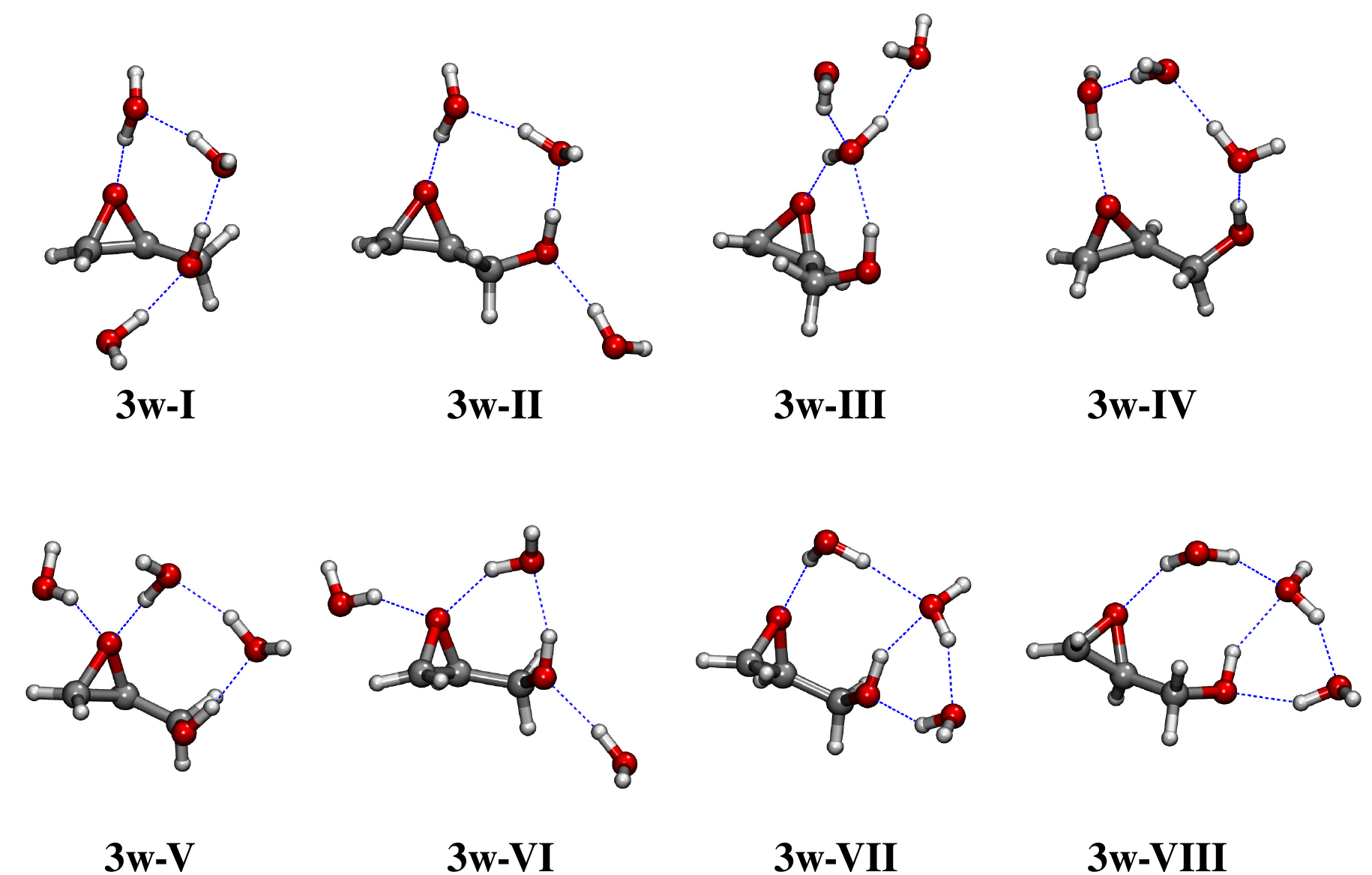}
\caption{Structures of the eight conformers of glycidol + 3 waters clusters. In MLDFT calculations, glycidol is active whereas water molecules are inactive.}
\label{fig:gly_3w}
\end{figure}

The MOXY is a rigid molecule, so the different solvated conformers mainly differs by the position of the water molecules. In this section we show how MLDFT can treat flexible solutes, and to this end we have selected glycidol (GLY), which is a derivative of MOXY where one hydrogen of the methyl group is replaced by the OH group (see Figs. \ref{fig:gly_1w}, \ref{fig:gly_2w}, \ref{fig:gly_3w}). In all MLDFT calculations, the GLY moiety is the active fragment and the water molecules are the inactive part. The presence of the hydroxyl group makes glycidol flexible up to the point that eight different conformers can be located in the gas phase potential energy surface (PES).\cite{yang2009probing,giovannini2018effective} %Glycidol exhibits a peculiar chiroptical response in aqueous solution, in particular in vibrational optical activity such as Vibrational Circular Dichroism (VCD).\cite{yang2009probing,giovannini2018effective} For this reason, it has been amply studied both theoretically and experimentally. 
%One of the approaches that have been investigated for the description of its VCD spectrum in aqueous solution was the so-called cluster method, similarly to the case of methyloxirane (see e.g. Ref. \cite{yang2009probing}). In this work, 

To build up a glycidol/water clusters, different structures constituted by GLY and one, two and three water molecules were constructed, by following the strategy reported in Ref. \citenum{yang2009probing}. Such structures are depicted in Figs. \ref{fig:gly_1w}, \ref{fig:gly_2w}, \ref{fig:gly_3w}. We note that the different structures not only differ by the position of the water molecules, but also by the conformation of glycidol. In particular, the six conformers constituted by GLY and one water (GLY+1w) are characterized by a different position of the water molecule. The latter interacts via hydrogen bonding with both the hydroxyl and epoxyl groups (\textbf{1w-I} and \textbf{1w-II}), with the epoxyl group only ( \textbf{1w-III} and \textbf{1w-IV}), or with only the oxygen atom of the hydroxyl group 
(\textbf{1w-V} and \textbf{1w-VI}). The inclusion of an additional water (GLY+2w) results in ten different conformers, which are shown in Fig. \ref{fig:gly_2w}. These contain three or four center bridges (conformers \textbf{2w-I}, \textbf{2w-II}, \textbf{2w-IV}, \textbf{2w-V}, \textbf{2w-VI}, \textbf{2w-VII} and \textbf{2w-VIII}) or are conformers where the two water molecules interact via hydrogen bonding with the epoxyl and hydroxyl groups (conformers \textbf{2w-III}, \textbf{2w-IX} and \textbf{2w-X}). If three explicit water molecules are added to GLY (GLY+3w), the conformational search provides eight main conformers, which are graphically depicted in Fig. \ref{fig:gly_3w}. Similarly to the previous case, some of them contain three or four center bridges (conformers \textbf{3w-I}, \textbf{3w-II}, \textbf{3w-V}, \textbf{3w-VI}), whereas in conformers \textbf{3w-IV}, \textbf{3w-VII} and \textbf{3w-VIII} a five center bridge is present. In all cases, water molecules that are not involved in bridges interact with GLY through hydrogen bonding interaction. Conformer \textbf{3w-III} is instead characterized by a three center bridge and by the remaining water molecules hydrogen bonded to the bridge water. 

\begin{figure}[htbp!]
\centering
\subfigure[][]
{\includegraphics[width=.33\textwidth]{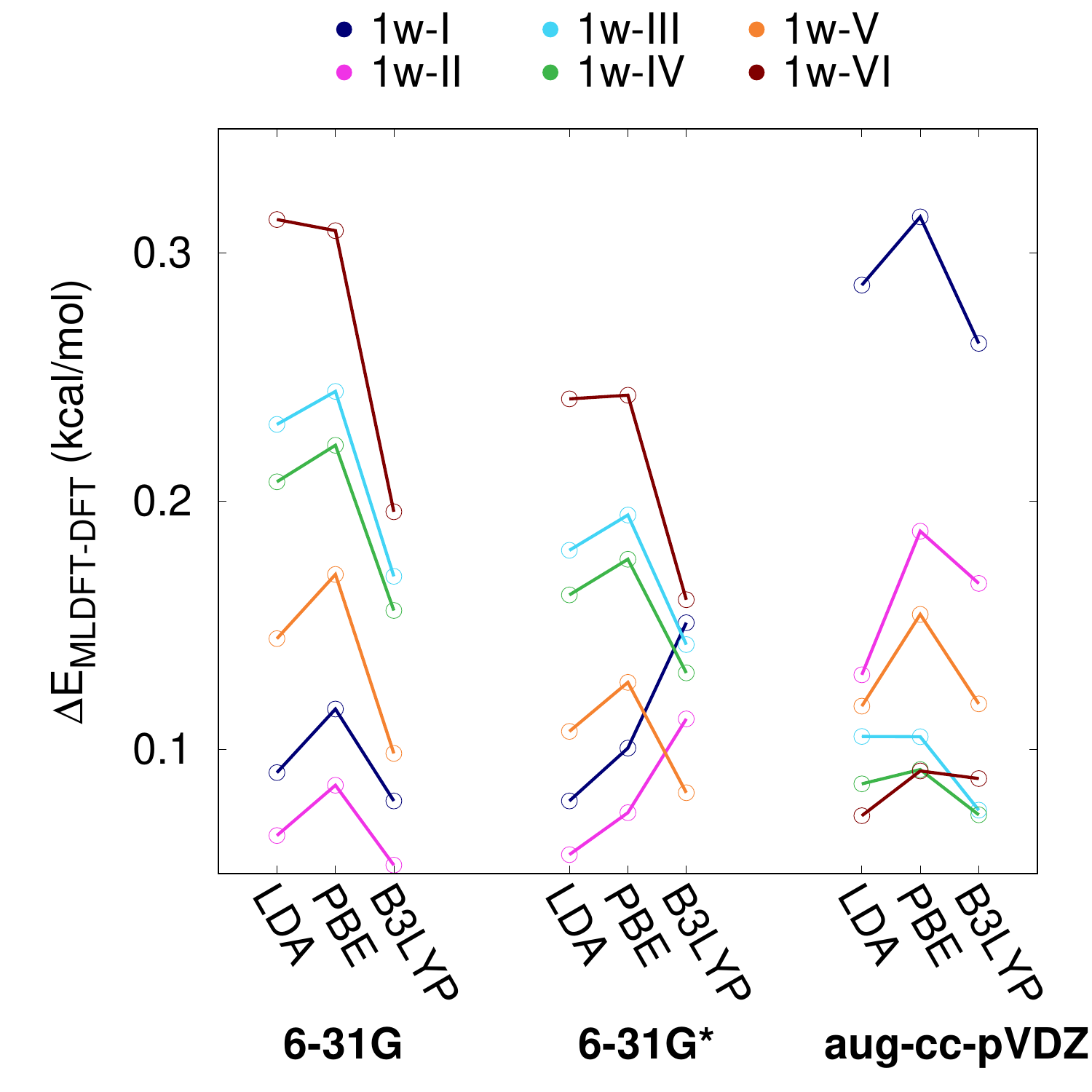}
\includegraphics[width=.33\textwidth]{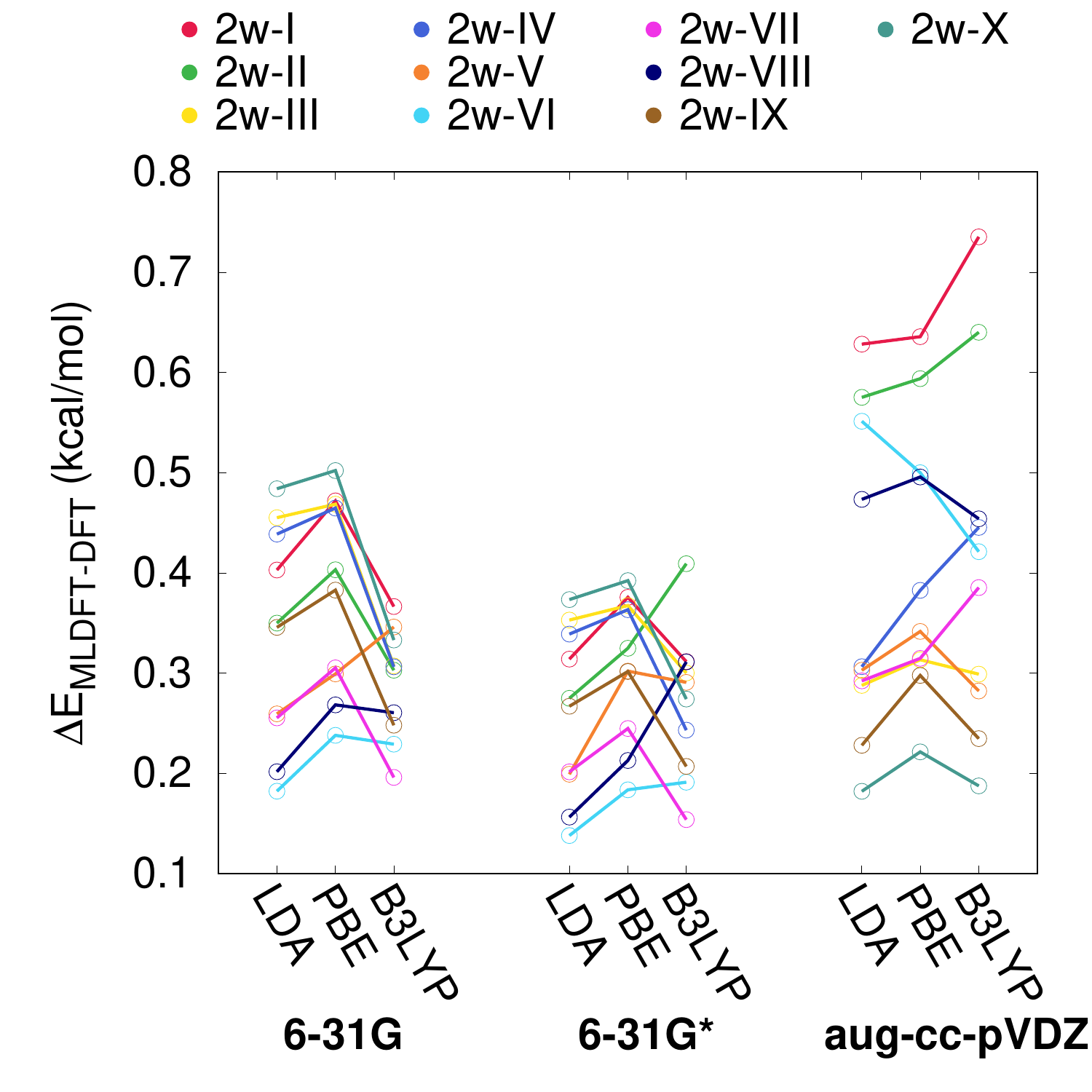}
\includegraphics[width=.33\textwidth]{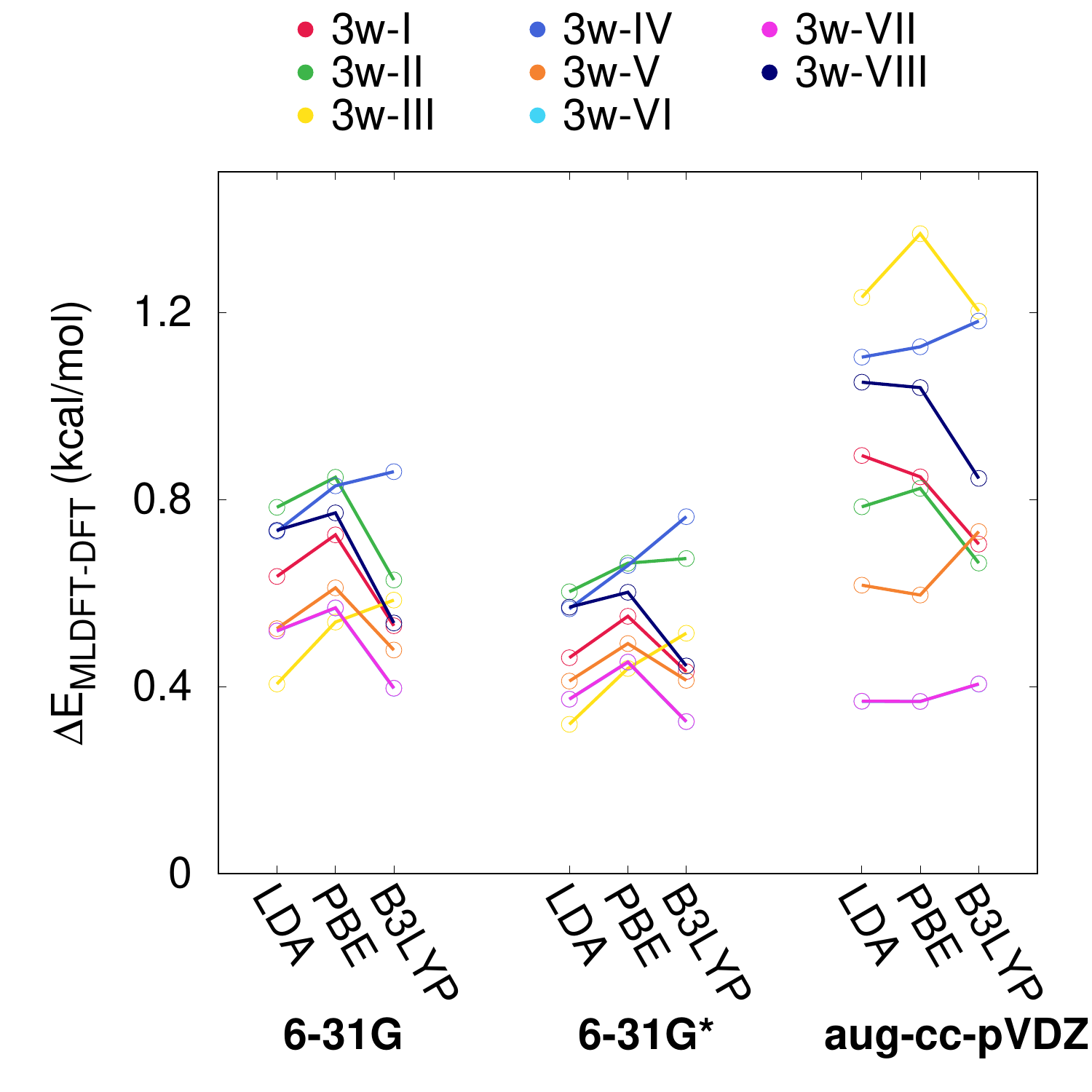}}
\subfigure[][]
{\includegraphics[width=.33\textwidth]{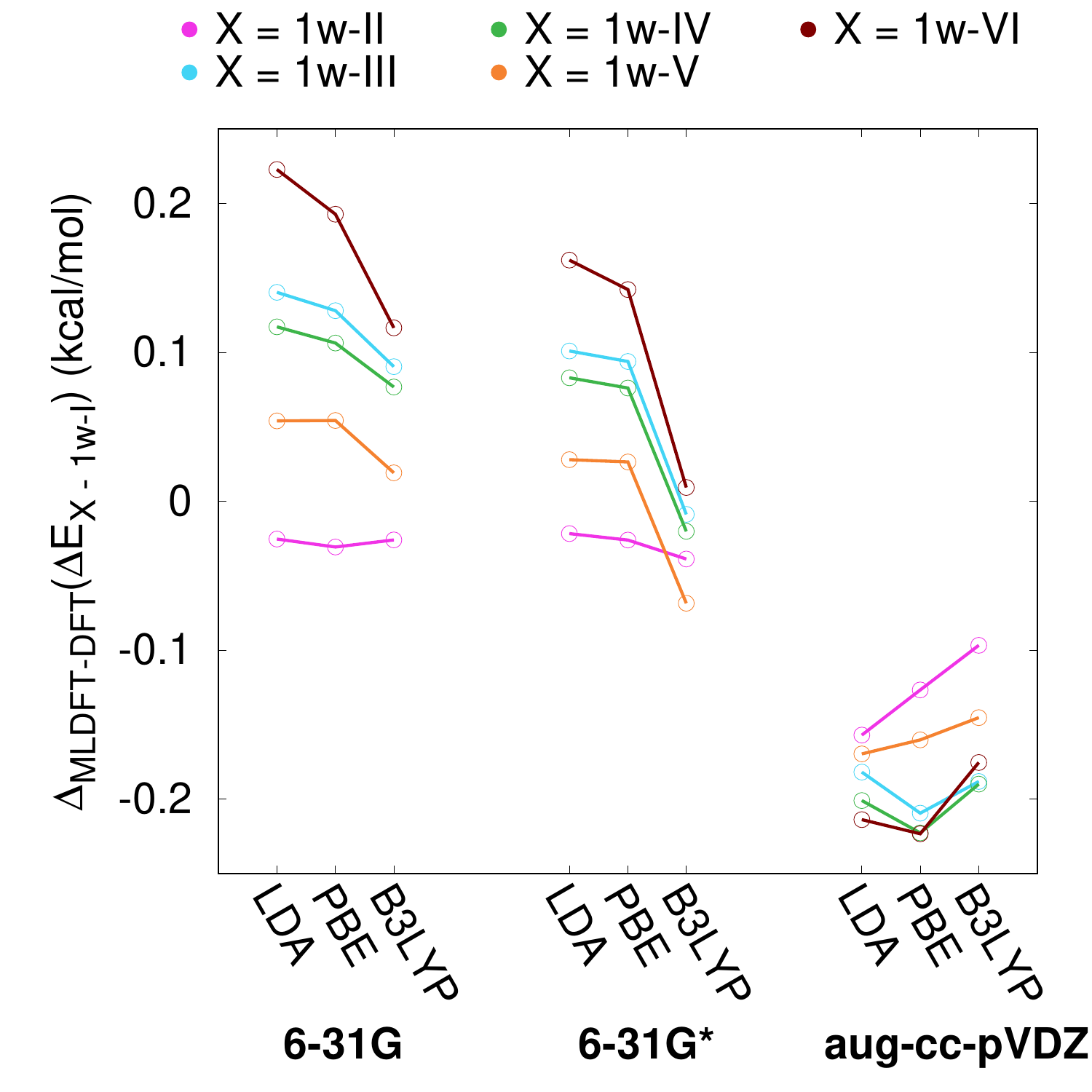}
\includegraphics[width=.33\textwidth]{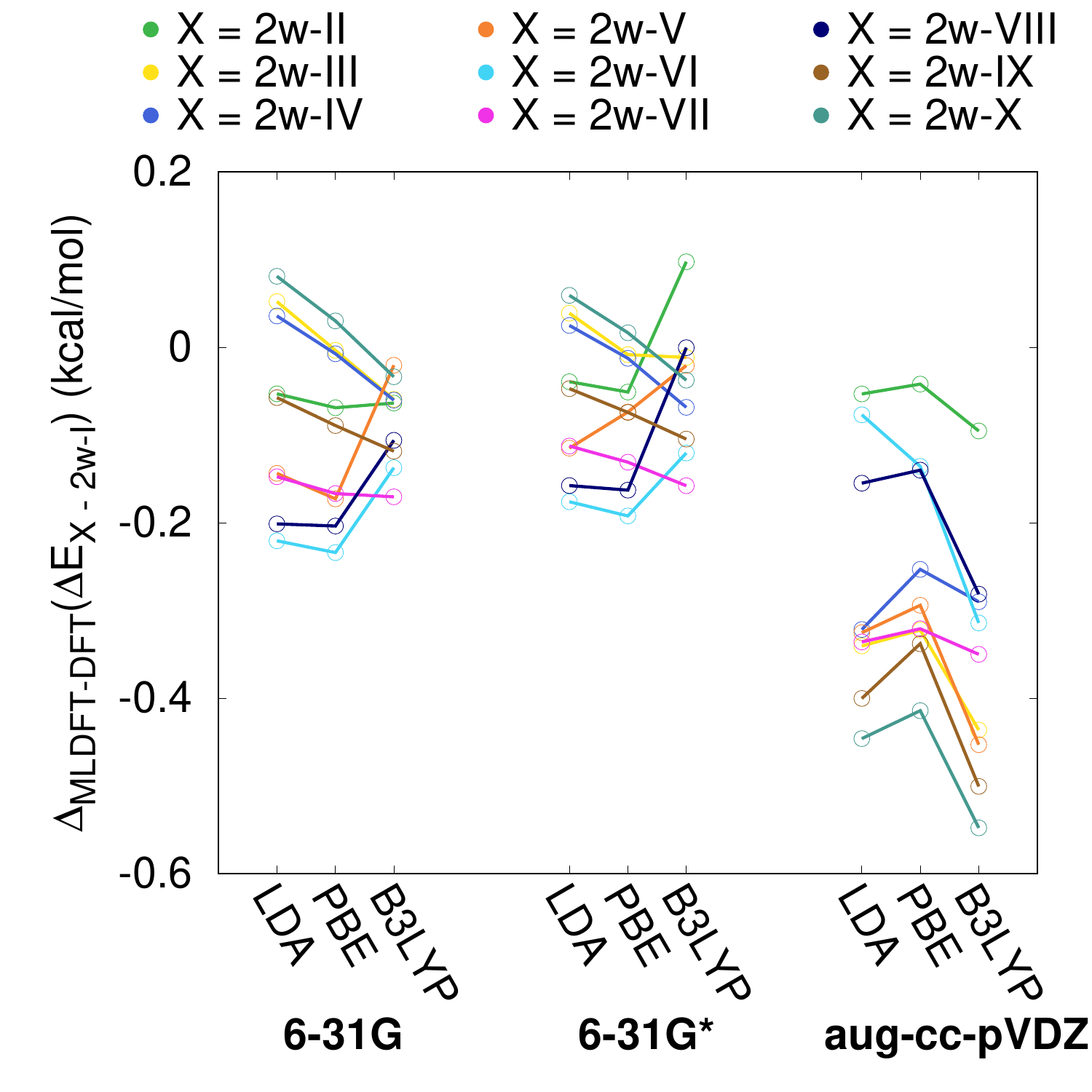}
\includegraphics[width=.33\textwidth]{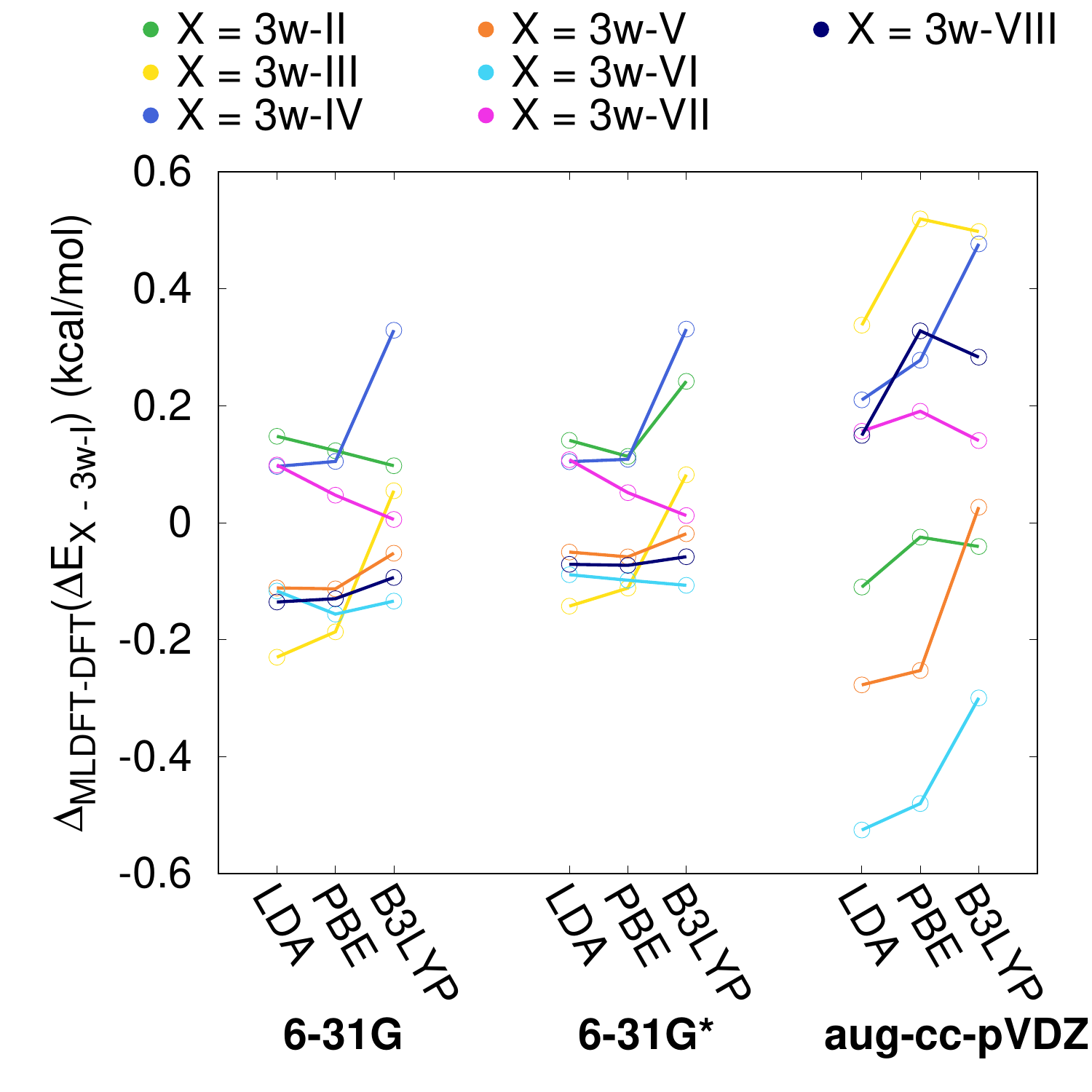}}
\caption{\textbf{(a)} GLY+1w (left), GLY+2w (middle) and GLY+3w (right) GS energy differences between MLDFT and reference DFT values. \textbf{(b)} MLDFT-DFT energy deviations for the energy differences between each conformer of GLY+1w conformers and \textbf{1w-I} (left), GLY+2w conformers and \textbf{2w-I} (middle) and GLY+3w conformers and \textbf{3w-I} (right). All values are reported in kcal/mol.}
\label{fig:gly_1w_2w_3w_data}
\end{figure}

We now move to discuss GS energy differences between DFT and MLDFT (see Fig. \ref{fig:gly_1w_2w_3w_data}a, raw data are given in Tabs. S3-S5 in the SI). 
%aggiustare la caption della figura con le energie. Riscrivere il commento su 1w, 2w e 3w insieme e in modo comparativo. 

%1W + 2W + 3W (GS energies) tommaso
In Fig. \ref{fig:gly_1w_2w_3w_data}, panel a, MLDFT - DFT GS energy differences for all the different conformers of GLY+1w, GLY+2w and GLY+3w water clusters are shown. The error reported by MLDFT is below 0.1 mH ($< 0.627$ kcal/mol) when applied to GLY+1W, at all the selected levels of theory. In particular, energy differences are perfectly in line with what is shown in Fig. \ref{fig:moxy_1w_2w_data}a, left panel, in case of MOXY+1w clusters. Moving to GLY+2w conformers, the agreement between DFT and MLDFT is almost perfect at all levels of theory, being the energy difference below 0.8 kcal/mol in all cases. We also see that at B3LYP/aug-cc-pVDZ level, for \textbf{2w-I} and \textbf{2w-II} the difference between MLDFT and full DFT is larger than for the other conformers ($> 0.1$ mH, 0.627 kcal/mol). This is due to the specific spatial arrangement of water molecules, which create a four-center bridge connecting GLY hydroxyl and epoxyl groups (see Fig. \ref{fig:gly_2w}). 

As stated above for MOXY+2w clusters, MLDFT accounts for all the interactions between active and inactive parts, with the exception of dispersion; however, the inactive fragment(s) are described by a frozen density. Therefore, polarization and charge transfer (and dispersion) effects are neglected in the inactive region. For \textbf{2w-I} and \textbf{2w-II} we can speculate that such interactions may play a relevant role, because the two inactive water molecules are hydrogen bonded. Also, their role is clearly increased when diffuse and polarization functions are included in the basis set (aug-cc-pVDZ), because such functions enhance the effects of these interactions. This is not occurring in case of the other conformers, because of the different spatial arrangement of the solvent molecules.

We now focus on GLY+3w conformers. The agreement between MLDFT and reference full DFT values is generally worse than in the previous cases (see right panel of Fig. \ref{fig:gly_1w_2w_3w_data}a). However, the average error is of about 0.67 kcal/mol ($\sim$ 0.1 mH), i.e. again well beyond the chemical accuracy. The largest discrepancy is shown by \textbf{3w-III} for all the functionals (LDA, PBE or B3LYP) in combination with aug-cc-pVDZ ($\sim$ 1.2 kcal/mol). Again, this can be explained by considering the spatial arrangement of water molecules around GLY 
(see Fig. \ref{fig:gly_3w}). Similar to \textbf{2w-I} and \textbf{2w-II}, the effect of charge transfer and polarization interactions, which are neglected by the partitioning of the inactive density in MLDFT, may play a relevant role. Such effects are larger for \textbf{3w-III}, however they affect also other conformers which are characterized by a four/five center bridge. It is also worth noticing that the MLDFT error is expected to increase with the size of the studied system, because the energy is an extensive quantity. Such a trend is in fact reported for both MOXY and GLY clusters.

Let us now discuss the MLDFT-DFT energy deviations for the energy differences between each conformer of the GLY clusters and \textbf{1w-I}, \textbf{2w-I} and \textbf{3w-I}, which are reported in Fig. \ref{fig:gly_1w_2w_3w_data}b. Raw data are given in Tabs. S3-S5 in the SI. 

For GLY+1w system, both MLDFT and DFT predict \textbf{1w-I} to be the most stable at all levels of theory, whereas the relative populations of the other conformers strongly depend on the theory level (see Fig. \ref{fig:gly_1w_2w_3w_data}b, left panel). In particular, the energy differences of each conformer with respect to \textbf{1w-I} decrease as larger basis sets are employed, and also by moving from LDA to PBE and B3LYP. The error between MLDFT and DFT is instead almost constant (in absolute value) for all different combinations of basis set and DFT functional, and in all cases MLDFT correctly reproduces the trends obtained at the reference full DFT level.

The same considerations outlined above for GLY+1w conformers, also apply to GLY+2w ones (see Fig. \ref{fig:gly_1w_2w_3w_data}b, middle panel). In fact, by moving from LDA to B3LYP and by including polarization and diffuse functions in the basis set, MLDFT errors with respect to DFT reference values decrease. The largest DFT-MLDFT discrepancy is reported for \textbf{2w-X} at the B3LYP/aug-cc-pVDZ level (-0.55 kcal/mol). This is due to the fact that the largest error is associated to the GS energy of the most stable conformer \textbf{2w-I} (see left panel of Fig. \ref{fig:gly_1w_2w_3w_data}a) for this combination of DFT functional/basis set. However, as already reported for all the other studied systems, the error in the relative energies of the different conformers is always lower than the corresponding error in the total energies.

Finally, also in case of GLY+3w clusters the agreement between DFT and MLDFT is almost perfect, with errors ranging from -0.6 to 0.6 kcal/mol. The maximum error is observed for \textbf{3w-III} at the PBE/aug-cc-pVDZ level (0.53 kcal/mol), whereas the minimum is reported for \textbf{3w-VII} at the B3LYP/6-31G* level (error $<$ 0.01 kcal/mol). Therefore, also for these systems, MLDFT provides a reliable description of the relative energies of the different conformers. The only notable exceptions are conformers \textbf{3w-III} and \textbf{3w-IV} at the LDA/6-31G and B3LYP/6-31G levels, respectively. As a final comment, we note that, although the MLDFT error on total ground state energy can be larger than 1 kcal/mol, relative energies of the different conformers are accurately predicted, with an error that is always below 1 kcal/mol.

\subsection{Towards a Realistic Picture of Solvation}

In the previous sections, we have presented and discussed solute-solvent structures obtained by modeling the solvation phenomenon in aqueous solution by means of the so-called cluster approach,\cite{perera2016clusters} in which only the closest water molecules are explicitly treated at the QM level. However, this picture is not realistic, being a strongly approximate way of modeling solvation. In fact, any dynamical aspect of solvation is neglected as well as, more importantly, long range interactions which are especially relevant for polar environments such as water. In this section, we show how MLDFT may be coupled to approaches that have been developed to model solvation more realistically. In particular, we will apply MLDFT to a randomly selected structure extracted from a classical MD simulation performed on both MOXY and GLY in aqueous solution. In this way, the atomistic details of solvation are retained, and dynamical aspects could easily be introduced by repeating the calculations on several structures. A closer investigation of the latter aspect is beyond the scope of our first work on MLDFT, and will be the topic of further studies.

Let us start with MOXY. We have selected one random snapshot extracted from a MD simulation, which was previously reported by some of the present authors.\cite{lipparini2013optical,giovannini2017polarizable,giovannini2016effective} Note that MOXY is a rigid molecule, therefore a single snapshot well represents its conformational structure.

\begin{figure}[htbp!]
\centering
\includegraphics[width=.48\textwidth]{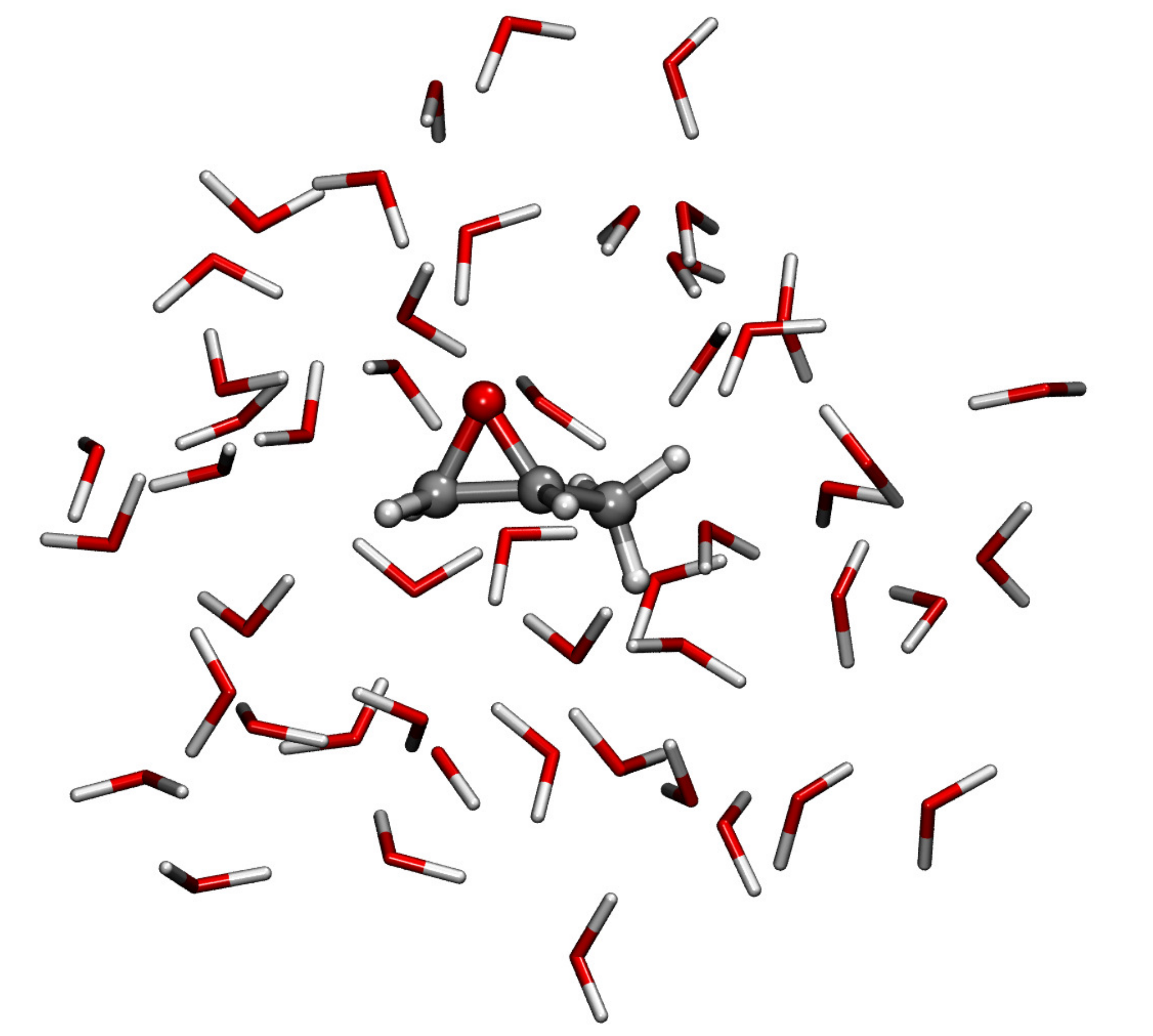}
\caption{Selected structure of MOXY + 50 water molecules, as extracted from MD. In MLDFT calculations, MOXY is the active part and water molecules are inactive.}
\label{fig:moxy_snap}
\end{figure}

In MLDFT calculations, MOXY is the active fragment and it is treated at the B3LYP/6-31+G* level. The inactive part is constituted by the 50 closest water molecules, which are described at the B3LYP/6-31G level. The reference full DFT calculation is instead performed by using the B3LYP functional, in combination with the 6-31+G* basis set for MOXY and the 6-31G one for water molecules.

In order to quantify the accuracy of MLDFT, we compute the solvation energy $E_{solv}$, which is defined as:

\begin{equation}
E_{\text{solv}} = E_{\text{tot}} - E_{\text{MOXY}} - E_{\text{w}}\ ,
\label{eq:solv-moxy}
\end{equation}

where $E_{\text{tot}}$, $E_{\text{MOXY}}$ and $E_\text{w}$ are the total, MOXY and water GS energies, respectively. Note that $E_{\text{MOXY}}$ is calculated in the gas phase, and thus it is the same in both full DFT and MLDFT calculations. The $E_{\text{tot}}$ and $E_{\text{w}}$ are defined differently in the two approaches; in MLDFT $E_{\text{w}}$ is calculated at step 1 of the computational protocol (see section 3), whereas in full DFT it refers to the GS energy of the 50 water molecules.

\begin{table}[htbp!]
\centering
\begin{tabular}[t]{lrr}
\hline
                            &  DFT & MLDFT \\
\hline
$E_{\text{tot}}$            & -4013.1956 & -4013.1660\\
$E_{\text{MOXY}}$           &  -193.1079 &  -193.1079 \\
$E_{\text{w}}$              & -3820.0681 & -3820.0382  \\
\hline
$E_{\text{solv}}$           & -0.0196   & -0.0199  \\
$E_{\text{solv}}$(kcal/mol) & -12.3014  & -12.4766 \\
\hline
\end{tabular}
\caption{DFT and MLDFT total GS energies ($E_{\text{tot}}$) of MOXY + 50 water molecules snapshot depicted in Fig. \ref{fig:moxy_snap}. $E_{\text{MOXY}}$, $E_\text{w}$ and $E_{\text{solv}}$ are also reported. All values are given in Hartree, unless specified.}
\label{tab:moxy_snap}
\end{table}

Computed energy values for MOXY are reported in Tab. \ref{tab:moxy_snap} for both DFT and MLDFT. We first notice that the MLDFT error on the total energy $E_{\text{tot}}$ is larger than what is found for clusters (see previous sections). This is not surprising, because the error of the method scales with the number of the water molecules in the inactive part. Such discrepancies are primarily due to the neglect of polarization and charge-transfer interactions in the inactive solvent water molecules, because their density remains fixed in MLDFT. The largest contribution to the error on total energy is due to $E_{\text{w}}$. In fact, MLDFT $E_{\text{w}}$ differs from full DFT of about the same extent as total energies. Such differences between MLDFT and DFT are reflected by the computed solvation energy, which can be taken as a measure of the accuracy of MLDFT. For the studied snapshot, the agreement between MLDFT and DFT is almost perfect, and the error is of about 0.2 kcal/mol.  

\begin{figure}[htbp!]
\centering
\includegraphics[width=.85\textwidth]{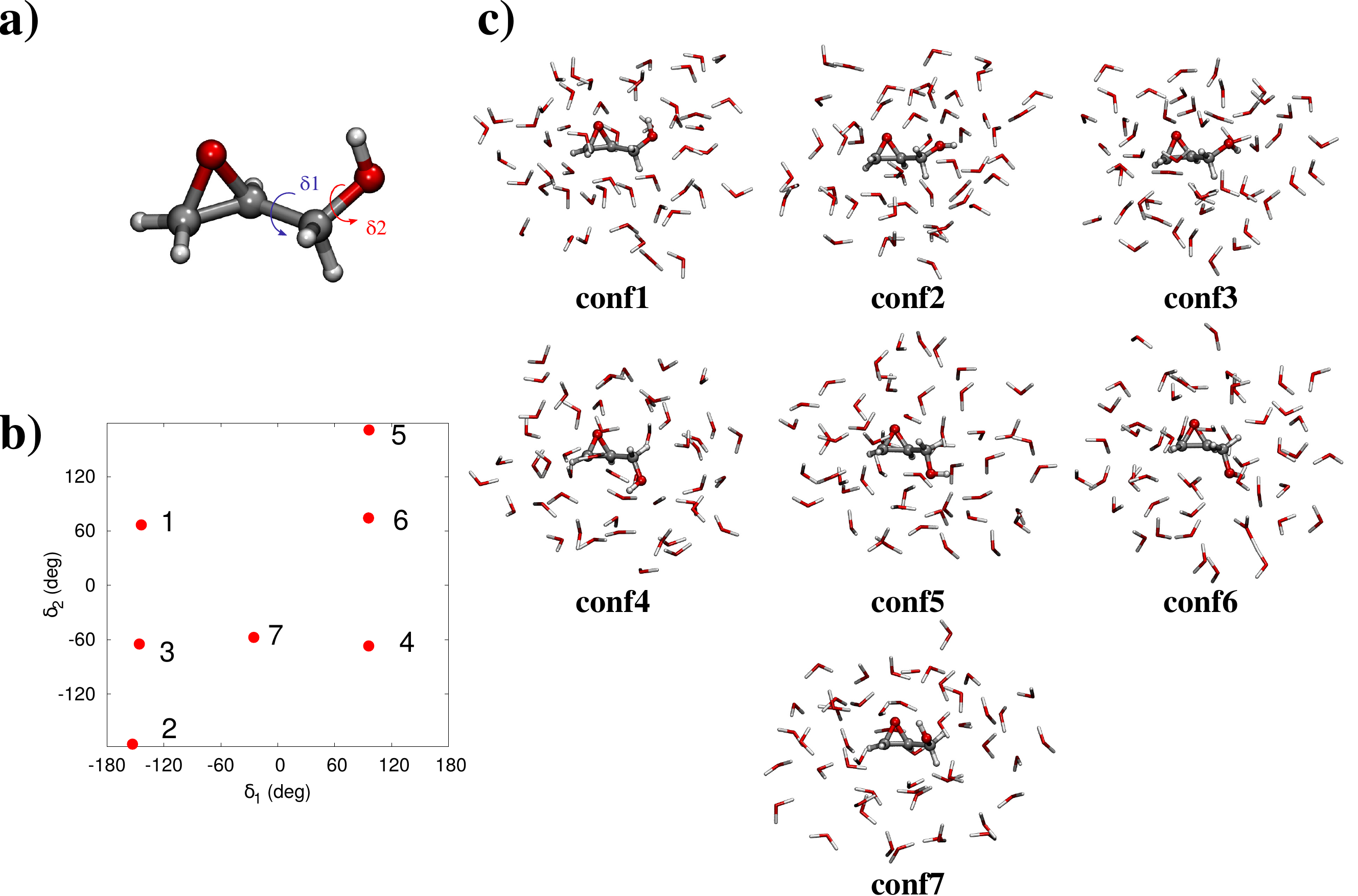}
\caption{a) Definition of $\delta_1$ and $\delta_2$ dihedral angles of GLY; b) $\delta_1$ and $\delta_2$ values for the selected GLY + 50 water molecules snaphosts extracted from MD; c) molecular structures of the seven selected snapshots. In MLDFT calculations, GLY is the active part, whereas water molecules are inactive.}
\label{fig:gly_snap}
\end{figure}

The same analysis may be applied to glycidol, for which the snapshots were extracted from MD simulations previously reported by some of us.\cite{giovannini2018effective} We recall that GLY is a flexible solute, of which the main conformers may be identified by means of two dihedral angles $\delta1$ and $\delta2$  (see Fig. \ref{fig:gly_snap}, panel a). Seven most probable conformers have been selected (see Fig. \ref{fig:gly_snap} panel b). 

The MLDFT partition has been done so that GLY is the active fragment, and treated at the B3LYP/6-31+G* level, whereas water molecules are inactive and described at the B3LYP/6-31G level. All the reference full DFT calculations are performed by using the B3LYP functional in combination with the 6-31+G* basis set for the solute and the 6-31G one for the water molecules. 

The DFT and MLDFT energies ($E_{\text{tot}}$, $E_{\text{GLY}}$, $E_{\text{w}}$ and $E_{\text{solv}}$) are reported in Tab. S6 in the SI. Overall, MLDFT total energies are higher than DFT values of about 0.02-0.03 Hartree. The reasons of this discrepancy is the same as reported for MOXY.

\begin{figure}[htbp!]
\centering
\includegraphics[width=.5\textwidth]{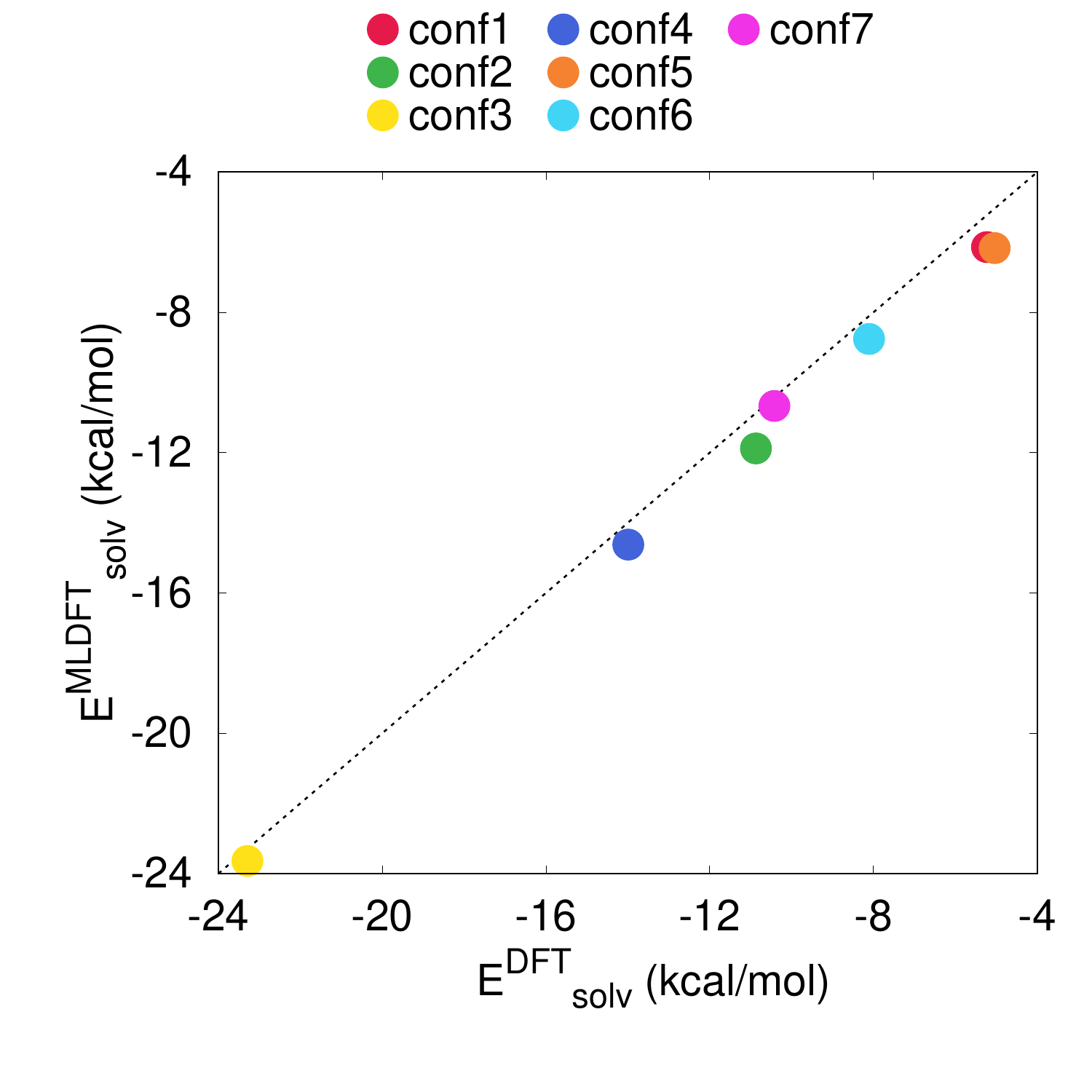}
\caption{DFT and MLDFT solvation energies ($E_{\text{solv}}$) for the different conformers graphically depicted in Fig. \ref{fig:gly_snap}. Values are given in kcal/mol. }
\label{fig:gly_snap_solv}
\end{figure}

The DFT and MLDFT solvation energies are graphically compared in Fig. \ref{fig:gly_snap_solv}. We observe that all MLDFT values are almost in perfect agreement with the reference full DFT data. The average discrepancy is of about 0.7 kcal/mol ($\sim$ 1 mH), with the largest discrepancy reported for conformer 5 (1.1 kcal/mol). Notice that in this study we only include the GLY moiety in the active part. Similar calculations performed at the MLHF level \cite{saether2017density} needed to insert at least 5 water molecules in the active fragment to reach the same level of accuracy. 

\section{Summary and Conclusions}

In this work, we report a novel density-based multilevel approach based on a DFT treatment of the electronic structure problem. In MLDFT, the studied system is partitioned in two layers, one active and one inactive. The unique characteristic of MLDFT is the SCF procedure, that is performed in the MO basis of the active part only. This allows for a reduction in computational cost, because the inactive fragments are kept frozen during the optimization of the density. 

The MLDFT was applied to aqueous methyloxirane and glycidol, for which two different approaches to solvation were discussed. First, the so-called cluster approach is employed, which models solvation in terms of minimal clusters composed of the solute and a small number of water molecules. Second, a more realistic picture is considered, which focuses on randomly selected snapshots extracted from MD simulations. For all studied structures, the computed data confirm that MLDFT is able to correctly reproduce reference full DFT values, with errors which are always $\le$ 1 kcal/mol.
Due to its favorable computational scaling, MLDFT can be coupled to more realistic approaches to solvation, i.e. it can treat a large number of representative snapshots extracted from MD simulations, so to effectively take into account the dynamical aspects of solvation. 

In this first presentation of the approach, we have limited the analysis to ground state energies. However, MLDFT has the potentialities to be extended to the calculation of molecular properties and spectra. Such extensions will be the topic of future communications.
 
The method will also be further developed by focusing on some technical aspects, which are worth being improved. For instance, in the current implementation, the DFT grid is homogeneous in the whole space. However, it is reasonable to assume that the grid can be downgraded further away from the active part. Technical refinements of the current implementation are in progress and will be discussed in future communications.

\section{Supporting Information}

Data related to Figs. \ref{fig:moxy_1w_2w_data}, \ref{fig:gly_1w_2w_3w_data} and \ref{fig:gly_snap_solv}.

\section{Acknowledgments}

\small{We acknowledge funding from the Marie Sklodowska-Curie European Training Network “COSINE - COmputational Spectroscopy In Natural sciences and Engineering”, Grant Agreement No.  765739, and the Research Council of Norway through FRINATEK projects 263110 and 275506.}

\newpage

{
\small
\bibliography{biblio}

\providecommand{\latin}[1]{#1}
\makeatletter
\providecommand{\doi}
  {\begingroup\let\do\@makeother\dospecials
  \catcode`\{=1 \catcode`\}=2 \doi@aux}
\providecommand{\doi@aux}[1]{\endgroup\texttt{#1}}
\makeatother
\providecommand*\mcitethebibliography{\thebibliography}
\csname @ifundefined\endcsname{endmcitethebibliography}
  {\let\endmcitethebibliography\endthebibliography}{}
\begin{mcitethebibliography}{82}
\providecommand*\natexlab[1]{#1}
\providecommand*\mciteSetBstSublistMode[1]{}
\providecommand*\mciteSetBstMaxWidthForm[2]{}
\providecommand*\mciteBstWouldAddEndPuncttrue
  {\def\EndOfBibitem{\unskip.}}
\providecommand*\mciteBstWouldAddEndPunctfalse
  {\let\EndOfBibitem\relax}
\providecommand*\mciteSetBstMidEndSepPunct[3]{}
\providecommand*\mciteSetBstSublistLabelBeginEnd[3]{}
\providecommand*\EndOfBibitem{}
\mciteSetBstSublistMode{f}
\mciteSetBstMaxWidthForm{subitem}{(\alph{mcitesubitemcount})}
\mciteSetBstSublistLabelBeginEnd
  {\mcitemaxwidthsubitemform\space}
  {\relax}
  {\relax}

\bibitem[Dykstra \latin{et~al.}(2011)Dykstra, Frenking, Kim, and
  Scuseria]{dykstra2011theory}
Dykstra,~C.; Frenking,~G.; Kim,~K.; Scuseria,~G. \emph{Theory and applications
  of computational chemistry: the first forty years}; Elsevier, 2011\relax
\mciteBstWouldAddEndPuncttrue
\mciteSetBstMidEndSepPunct{\mcitedefaultmidpunct}
{\mcitedefaultendpunct}{\mcitedefaultseppunct}\relax
\EndOfBibitem
\bibitem[Reichardt(1992)]{reichardt1992solvatochromism}
Reichardt,~C. Solvatochromism, thermochromism, piezochromism, halochromism, and
  chiro-solvatochromism of pyridinium N-phenoxide betaine dyes. \emph{Chem.
  Soc. Rev.} \textbf{1992}, \emph{21}, 147--153\relax
\mciteBstWouldAddEndPuncttrue
\mciteSetBstMidEndSepPunct{\mcitedefaultmidpunct}
{\mcitedefaultendpunct}{\mcitedefaultseppunct}\relax
\EndOfBibitem
\bibitem[Buncel and Rajagopal(1990)Buncel, and
  Rajagopal]{buncel1990solvatochromism}
Buncel,~E.; Rajagopal,~S. Solvatochromism and solvent polarity scales.
  \emph{Acc. Chem. Res.} \textbf{1990}, \emph{23}, 226--231\relax
\mciteBstWouldAddEndPuncttrue
\mciteSetBstMidEndSepPunct{\mcitedefaultmidpunct}
{\mcitedefaultendpunct}{\mcitedefaultseppunct}\relax
\EndOfBibitem
\bibitem[Reichardt(1994)]{reichardt1994solvatochromic}
Reichardt,~C. Solvatochromic dyes as solvent polarity indicators. \emph{Chem.
  Rev.} \textbf{1994}, \emph{94}, 2319--2358\relax
\mciteBstWouldAddEndPuncttrue
\mciteSetBstMidEndSepPunct{\mcitedefaultmidpunct}
{\mcitedefaultendpunct}{\mcitedefaultseppunct}\relax
\EndOfBibitem
\bibitem[Cannelli \latin{et~al.}(2017)Cannelli, Giovannini, Baiardi, Carlotti,
  Elisei, and Cappelli]{cannelli2017understanding}
Cannelli,~O.; Giovannini,~T.; Baiardi,~A.; Carlotti,~B.; Elisei,~F.;
  Cappelli,~C. Understanding the interplay between the solvent and nuclear
  rearrangements in the negative solvatochromism of a push--pull flexible
  quinolinium cation. \emph{Phys. Chem. Chem. Phys.} \textbf{2017}, \emph{19},
  32544--32555\relax
\mciteBstWouldAddEndPuncttrue
\mciteSetBstMidEndSepPunct{\mcitedefaultmidpunct}
{\mcitedefaultendpunct}{\mcitedefaultseppunct}\relax
\EndOfBibitem
\bibitem[Carlotti \latin{et~al.}(2018)Carlotti, Cesaretti, Cannelli,
  Giovannini, Cappelli, Bonaccorso, Fortuna, Elisei, and
  Spalletti]{carlotti2018evaluation}
Carlotti,~B.; Cesaretti,~A.; Cannelli,~O.; Giovannini,~T.; Cappelli,~C.;
  Bonaccorso,~C.; Fortuna,~C.~G.; Elisei,~F.; Spalletti,~A. Evaluation of
  Hyperpolarizability from the Solvatochromic Method: Thiophene Containing
  Push- Pull Cationic Dyes as a Case Study. \emph{J. Phys. Chem. C}
  \textbf{2018}, \emph{122}, 2285--2296\relax
\mciteBstWouldAddEndPuncttrue
\mciteSetBstMidEndSepPunct{\mcitedefaultmidpunct}
{\mcitedefaultendpunct}{\mcitedefaultseppunct}\relax
\EndOfBibitem
\bibitem[Cupellini \latin{et~al.}(2020)Cupellini, Calvani, Jacquemin, and
  Mennucci]{cupellini2020charge}
Cupellini,~L.; Calvani,~D.; Jacquemin,~D.; Mennucci,~B. Charge transfer from
  the carotenoid can quench chlorophyll excitation in antenna complexes of
  plants. \emph{Nat. Commun.} \textbf{2020}, \emph{11}, 1--8\relax
\mciteBstWouldAddEndPuncttrue
\mciteSetBstMidEndSepPunct{\mcitedefaultmidpunct}
{\mcitedefaultendpunct}{\mcitedefaultseppunct}\relax
\EndOfBibitem
\bibitem[Bondanza \latin{et~al.}(2020)Bondanza, Cupellini, Lipparini, and
  Mennucci]{bondanza2020multiple}
Bondanza,~M.; Cupellini,~L.; Lipparini,~F.; Mennucci,~B. The multiple roles of
  the protein in the photoactivation of Orange Carotenoid Protein. \emph{Chem}
  \textbf{2020}, \emph{6}, 187--203\relax
\mciteBstWouldAddEndPuncttrue
\mciteSetBstMidEndSepPunct{\mcitedefaultmidpunct}
{\mcitedefaultendpunct}{\mcitedefaultseppunct}\relax
\EndOfBibitem
\bibitem[Lomize \latin{et~al.}(2006)Lomize, Pogozheva, Lomize, and
  Mosberg]{lomize2006positioning}
Lomize,~A.~L.; Pogozheva,~I.~D.; Lomize,~M.~A.; Mosberg,~H.~I. Positioning of
  proteins in membranes: a computational approach. \emph{Protein Sci.}
  \textbf{2006}, \emph{15}, 1318--1333\relax
\mciteBstWouldAddEndPuncttrue
\mciteSetBstMidEndSepPunct{\mcitedefaultmidpunct}
{\mcitedefaultendpunct}{\mcitedefaultseppunct}\relax
\EndOfBibitem
\bibitem[Furse and Corcelli(2008)Furse, and Corcelli]{furse2008dynamics}
Furse,~K.~E.; Corcelli,~S.~A. The dynamics of water at DNA interfaces:
  Computational studies of Hoechst 33258 bound to DNA. \emph{J. Am. Chem. Soc.}
  \textbf{2008}, \emph{130}, 13103--13109\relax
\mciteBstWouldAddEndPuncttrue
\mciteSetBstMidEndSepPunct{\mcitedefaultmidpunct}
{\mcitedefaultendpunct}{\mcitedefaultseppunct}\relax
\EndOfBibitem
\bibitem[Zhao and Wei(2004)Zhao, and Wei]{zhao2004high}
Zhao,~S.; Wei,~G. High-order FDTD methods via derivative matching for Maxwell's
  equations with material interfaces. \emph{J. Comput. Phys.} \textbf{2004},
  \emph{200}, 60--103\relax
\mciteBstWouldAddEndPuncttrue
\mciteSetBstMidEndSepPunct{\mcitedefaultmidpunct}
{\mcitedefaultendpunct}{\mcitedefaultseppunct}\relax
\EndOfBibitem
\bibitem[Myhre and Koch(2016)Myhre, and Koch]{myhre2016multilevel}
Myhre,~R.~H.; Koch,~H. The multilevel CC3 coupled cluster model. \emph{J. Chem.
  Phys.} \textbf{2016}, \emph{145}, 044111\relax
\mciteBstWouldAddEndPuncttrue
\mciteSetBstMidEndSepPunct{\mcitedefaultmidpunct}
{\mcitedefaultendpunct}{\mcitedefaultseppunct}\relax
\EndOfBibitem
\bibitem[Høyvik \latin{et~al.}(2017)Høyvik, Myhre, and
  Koch]{hoyvik2017correlated}
Høyvik,~I.-M.; Myhre,~R.~H.; Koch,~H. Correlated natural transition orbitals
  for core excitation energies in multilevel coupled cluster models. \emph{J.
  Chem. Phys.} \textbf{2017}, \emph{146}, 144109\relax
\mciteBstWouldAddEndPuncttrue
\mciteSetBstMidEndSepPunct{\mcitedefaultmidpunct}
{\mcitedefaultendpunct}{\mcitedefaultseppunct}\relax
\EndOfBibitem
\bibitem[Parr(1980)]{parr1980density}
Parr,~R.~G. \emph{Horizons of quantum chemistry}; Springer, 1980; pp
  5--15\relax
\mciteBstWouldAddEndPuncttrue
\mciteSetBstMidEndSepPunct{\mcitedefaultmidpunct}
{\mcitedefaultendpunct}{\mcitedefaultseppunct}\relax
\EndOfBibitem
\bibitem[Burke(2012)]{burke2012perspective}
Burke,~K. Perspective on density functional theory. \emph{J. Chem. Phys.}
  \textbf{2012}, \emph{136}, 150901\relax
\mciteBstWouldAddEndPuncttrue
\mciteSetBstMidEndSepPunct{\mcitedefaultmidpunct}
{\mcitedefaultendpunct}{\mcitedefaultseppunct}\relax
\EndOfBibitem
\bibitem[Sisto \latin{et~al.}(2017)Sisto, Stross, van~der Kamp, O’Connor,
  McIntosh-Smith, Johnson, Hohenstein, Manby, Glowacki, and
  Martinez]{sisto2017atomistic}
Sisto,~A.; Stross,~C.; van~der Kamp,~M.~W.; O’Connor,~M.; McIntosh-Smith,~S.;
  Johnson,~G.~T.; Hohenstein,~E.~G.; Manby,~F.~R.; Glowacki,~D.~R.;
  Martinez,~T.~J. Atomistic non-adiabatic dynamics of the LH2 complex with a
  GPU-accelerated ab initio exciton model. \emph{Phys. Chem. Chem. Phys.}
  \textbf{2017}, \emph{19}, 14924--14936\relax
\mciteBstWouldAddEndPuncttrue
\mciteSetBstMidEndSepPunct{\mcitedefaultmidpunct}
{\mcitedefaultendpunct}{\mcitedefaultseppunct}\relax
\EndOfBibitem
\bibitem[Tomasi \latin{et~al.}(2005)Tomasi, Mennucci, and Cammi]{tomasi2005}
Tomasi,~J.; Mennucci,~B.; Cammi,~R. Quantum mechanical continuum solvation
  models. \emph{Chem. Rev.} \textbf{2005}, \emph{105}, 2999--3094\relax
\mciteBstWouldAddEndPuncttrue
\mciteSetBstMidEndSepPunct{\mcitedefaultmidpunct}
{\mcitedefaultendpunct}{\mcitedefaultseppunct}\relax
\EndOfBibitem
\bibitem[Mennucci(2012)]{Mennucci12_386}
Mennucci,~B. Polarizable Continuum Model. \emph{WIREs Comput. Mol. Sci.}
  \textbf{2012}, \emph{2}, 386--404\relax
\mciteBstWouldAddEndPuncttrue
\mciteSetBstMidEndSepPunct{\mcitedefaultmidpunct}
{\mcitedefaultendpunct}{\mcitedefaultseppunct}\relax
\EndOfBibitem
\bibitem[Tomasi and Persico(1994)Tomasi, and Persico]{tomasi1994molecular}
Tomasi,~J.; Persico,~M. Molecular interactions in solution: an overview of
  methods based on continuous distributions of the solvent. \emph{Chem. Rev.}
  \textbf{1994}, \emph{94}, 2027--2094\relax
\mciteBstWouldAddEndPuncttrue
\mciteSetBstMidEndSepPunct{\mcitedefaultmidpunct}
{\mcitedefaultendpunct}{\mcitedefaultseppunct}\relax
\EndOfBibitem
\bibitem[Cappelli(2016)]{cappelli2016integrated}
Cappelli,~C. Integrated QM/Polarizable MM/Continuum Approaches to Model
  Chiroptical Properties of Strongly Interacting Solute-Solvent Systems.
  \emph{Int. J. Quantum Chem.} \textbf{2016}, \emph{116}, 1532--1542\relax
\mciteBstWouldAddEndPuncttrue
\mciteSetBstMidEndSepPunct{\mcitedefaultmidpunct}
{\mcitedefaultendpunct}{\mcitedefaultseppunct}\relax
\EndOfBibitem
\bibitem[Warshel and Karplus(1972)Warshel, and Karplus]{warshel1972calculation}
Warshel,~A.; Karplus,~M. Calculation of ground and excited state potential
  surfaces of conjugated molecules. I. Formulation and parametrization.
  \emph{J. Am. Chem. Soc.} \textbf{1972}, \emph{94}, 5612--5625\relax
\mciteBstWouldAddEndPuncttrue
\mciteSetBstMidEndSepPunct{\mcitedefaultmidpunct}
{\mcitedefaultendpunct}{\mcitedefaultseppunct}\relax
\EndOfBibitem
\bibitem[Warshel and Levitt(1976)Warshel, and Levitt]{warshel1976theoretical}
Warshel,~A.; Levitt,~M. Theoretical studies of enzymic reactions: dielectric,
  electrostatic and steric stabilization of the carbonium ion in the reaction
  of lysozyme. \emph{J. Mol. Biol.} \textbf{1976}, \emph{103}, 227--249\relax
\mciteBstWouldAddEndPuncttrue
\mciteSetBstMidEndSepPunct{\mcitedefaultmidpunct}
{\mcitedefaultendpunct}{\mcitedefaultseppunct}\relax
\EndOfBibitem
\bibitem[Senn and Thiel(2009)Senn, and Thiel]{senn2009qm}
Senn,~H.~M.; Thiel,~W. {QM/MM} methods for biomolecular systems. \emph{Angew.
  Chem. Int. Ed.} \textbf{2009}, \emph{48}, 1198--1229\relax
\mciteBstWouldAddEndPuncttrue
\mciteSetBstMidEndSepPunct{\mcitedefaultmidpunct}
{\mcitedefaultendpunct}{\mcitedefaultseppunct}\relax
\EndOfBibitem
\bibitem[Lin and Truhlar(2007)Lin, and Truhlar]{lin2007qm}
Lin,~H.; Truhlar,~D.~G. {QM/MM}: what have we learned, where are we, and where
  do we go from here? \emph{Theor. Chem. Acc.} \textbf{2007}, \emph{117},
  185--199\relax
\mciteBstWouldAddEndPuncttrue
\mciteSetBstMidEndSepPunct{\mcitedefaultmidpunct}
{\mcitedefaultendpunct}{\mcitedefaultseppunct}\relax
\EndOfBibitem
\bibitem[Curutchet \latin{et~al.}(2009)Curutchet, Mu{\~n}oz-Losa, Monti,
  Kongsted, Scholes, and Mennucci]{curutchet2009electronic}
Curutchet,~C.; Mu{\~n}oz-Losa,~A.; Monti,~S.; Kongsted,~J.; Scholes,~G.~D.;
  Mennucci,~B. Electronic energy transfer in condensed phase studied by a
  polarizable QM/MM model. \emph{J. Chem. Theory Comput.} \textbf{2009},
  \emph{5}, 1838--1848\relax
\mciteBstWouldAddEndPuncttrue
\mciteSetBstMidEndSepPunct{\mcitedefaultmidpunct}
{\mcitedefaultendpunct}{\mcitedefaultseppunct}\relax
\EndOfBibitem
\bibitem[Olsen and Kongsted(2011)Olsen, and Kongsted]{olsen2011molecular}
Olsen,~J. M.~H.; Kongsted,~J. Molecular properties through polarizable
  embedding. \emph{Adv. Quantum Chem.} \textbf{2011}, \emph{61}, 107--143\relax
\mciteBstWouldAddEndPuncttrue
\mciteSetBstMidEndSepPunct{\mcitedefaultmidpunct}
{\mcitedefaultendpunct}{\mcitedefaultseppunct}\relax
\EndOfBibitem
\bibitem[Giovannini \latin{et~al.}(2019)Giovannini, Puglisi, Ambrosetti, and
  Cappelli]{giovannini2019fqfmu}
Giovannini,~T.; Puglisi,~A.; Ambrosetti,~M.; Cappelli,~C. Polarizable QM/MM
  approach with fluctuating charges and fluctuating dipoles: the QM/FQF$\mu$
  model. \emph{J. Chem. Theory Comput.} \textbf{2019}, \emph{15},
  2233--2245\relax
\mciteBstWouldAddEndPuncttrue
\mciteSetBstMidEndSepPunct{\mcitedefaultmidpunct}
{\mcitedefaultendpunct}{\mcitedefaultseppunct}\relax
\EndOfBibitem
\bibitem[Giovannini \latin{et~al.}(2017)Giovannini, Lafiosca, and
  Cappelli]{giovannini2017disrep}
Giovannini,~T.; Lafiosca,~P.; Cappelli,~C. A General Route to Include Pauli
  Repulsion and Quantum Dispersion Effects in QM/MM Approaches. \emph{J. Chem.
  Theory Comput.} \textbf{2017}, \emph{13}, 4854--4870\relax
\mciteBstWouldAddEndPuncttrue
\mciteSetBstMidEndSepPunct{\mcitedefaultmidpunct}
{\mcitedefaultendpunct}{\mcitedefaultseppunct}\relax
\EndOfBibitem
\bibitem[Giovannini \latin{et~al.}(2019)Giovannini, Lafiosca, Chandramouli,
  Barone, and Cappelli]{giovannini2019eprdisrep}
Giovannini,~T.; Lafiosca,~P.; Chandramouli,~B.; Barone,~V.; Cappelli,~C.
  Effective yet Reliable Computation of Hyperfine Coupling Constants in
  Solution by a QM/MM Approach: Interplay Between Electrostatics and
  Non-electrostatic Effects. \emph{J. Chem. Phys.} \textbf{2019}, \emph{150},
  124102\relax
\mciteBstWouldAddEndPuncttrue
\mciteSetBstMidEndSepPunct{\mcitedefaultmidpunct}
{\mcitedefaultendpunct}{\mcitedefaultseppunct}\relax
\EndOfBibitem
\bibitem[Giovannini \latin{et~al.}(2019)Giovannini, Ambrosetti, and
  Cappelli]{giovannini2019quantum}
Giovannini,~T.; Ambrosetti,~M.; Cappelli,~C. Quantum Confinement Effects on
  Solvatochromic Shifts of Molecular Solutes. \emph{J. Phys. Chem. Lett.}
  \textbf{2019}, \emph{10}, 5823--5829\relax
\mciteBstWouldAddEndPuncttrue
\mciteSetBstMidEndSepPunct{\mcitedefaultmidpunct}
{\mcitedefaultendpunct}{\mcitedefaultseppunct}\relax
\EndOfBibitem
\bibitem[Gordon \latin{et~al.}(2013)Gordon, Smith, Xu, and
  Slipchenko]{gordon2013accurate}
Gordon,~M.~S.; Smith,~Q.~A.; Xu,~P.; Slipchenko,~L.~V. Accurate first
  principles model potentials for intermolecular interactions. \emph{Annu. Rev.
  Phys. Chem.} \textbf{2013}, \emph{64}, 553--578\relax
\mciteBstWouldAddEndPuncttrue
\mciteSetBstMidEndSepPunct{\mcitedefaultmidpunct}
{\mcitedefaultendpunct}{\mcitedefaultseppunct}\relax
\EndOfBibitem
\bibitem[Gordon \latin{et~al.}(2007)Gordon, Slipchenko, Li, and
  Jensen]{gordon2007effective}
Gordon,~M.~S.; Slipchenko,~L.; Li,~H.; Jensen,~J.~H. The effective fragment
  potential: a general method for predicting intermolecular interactions.
  \emph{Annu. Rep. Comput. Chem.} \textbf{2007}, \emph{3}, 177--193\relax
\mciteBstWouldAddEndPuncttrue
\mciteSetBstMidEndSepPunct{\mcitedefaultmidpunct}
{\mcitedefaultendpunct}{\mcitedefaultseppunct}\relax
\EndOfBibitem
\bibitem[Sun and Chan(2016)Sun, and Chan]{sun2016quantum}
Sun,~Q.; Chan,~G. K.-L. Quantum embedding theories. \emph{Acc. Chem. Res.}
  \textbf{2016}, \emph{49}, 2705--2712\relax
\mciteBstWouldAddEndPuncttrue
\mciteSetBstMidEndSepPunct{\mcitedefaultmidpunct}
{\mcitedefaultendpunct}{\mcitedefaultseppunct}\relax
\EndOfBibitem
\bibitem[Knizia and Chan(2013)Knizia, and Chan]{knizia2013density}
Knizia,~G.; Chan,~G. K.-L. Density matrix embedding: A strong-coupling quantum
  embedding theory. \emph{J. Chem. Theory. Comput.} \textbf{2013}, \emph{9},
  1428--1432\relax
\mciteBstWouldAddEndPuncttrue
\mciteSetBstMidEndSepPunct{\mcitedefaultmidpunct}
{\mcitedefaultendpunct}{\mcitedefaultseppunct}\relax
\EndOfBibitem
\bibitem[Chulhai and Goodpaster(2018)Chulhai, and
  Goodpaster]{chulhai2018projection}
Chulhai,~D.~V.; Goodpaster,~J.~D. Projection-based correlated wave function in
  density functional theory embedding for periodic systems. \emph{J. Chem.
  Theory Comput.} \textbf{2018}, \emph{14}, 1928--1942\relax
\mciteBstWouldAddEndPuncttrue
\mciteSetBstMidEndSepPunct{\mcitedefaultmidpunct}
{\mcitedefaultendpunct}{\mcitedefaultseppunct}\relax
\EndOfBibitem
\bibitem[Chulhai and Goodpaster(2017)Chulhai, and
  Goodpaster]{chulhai2017improved}
Chulhai,~D.~V.; Goodpaster,~J.~D. Improved accuracy and efficiency in quantum
  embedding through absolute localization. \emph{J. Chem. Theory Comput.}
  \textbf{2017}, \emph{13}, 1503--1508\relax
\mciteBstWouldAddEndPuncttrue
\mciteSetBstMidEndSepPunct{\mcitedefaultmidpunct}
{\mcitedefaultendpunct}{\mcitedefaultseppunct}\relax
\EndOfBibitem
\bibitem[Wen \latin{et~al.}(2020)Wen, Graham, Chulhai, and
  Goodpaster]{wen2019absolutely}
Wen,~X.; Graham,~D.~S.; Chulhai,~D.~V.; Goodpaster,~J.~D. Absolutely Localized
  Projection-Based Embedding for Excited States. \emph{J. Chem. Theory Comput.}
  \textbf{2020}, \emph{16}, 385--398\relax
\mciteBstWouldAddEndPuncttrue
\mciteSetBstMidEndSepPunct{\mcitedefaultmidpunct}
{\mcitedefaultendpunct}{\mcitedefaultseppunct}\relax
\EndOfBibitem
\bibitem[Ding \latin{et~al.}(2017)Ding, Manby, and
  Miller~III]{ding2017embedded}
Ding,~F.; Manby,~F.~R.; Miller~III,~T.~F. Embedded mean-field theory with
  block-orthogonalized partitioning. \emph{J. Chem. Theory Comput.}
  \textbf{2017}, \emph{13}, 1605--1615\relax
\mciteBstWouldAddEndPuncttrue
\mciteSetBstMidEndSepPunct{\mcitedefaultmidpunct}
{\mcitedefaultendpunct}{\mcitedefaultseppunct}\relax
\EndOfBibitem
\bibitem[Goodpaster \latin{et~al.}(2012)Goodpaster, Barnes, Manby, and
  Miller~III]{goodpaster2012density}
Goodpaster,~J.~D.; Barnes,~T.~A.; Manby,~F.~R.; Miller~III,~T.~F. Density
  functional theory embedding for correlated wavefunctions: Improved methods
  for open-shell systems and transition metal complexes. \emph{J. Chem. Phys.}
  \textbf{2012}, \emph{137}, 224113\relax
\mciteBstWouldAddEndPuncttrue
\mciteSetBstMidEndSepPunct{\mcitedefaultmidpunct}
{\mcitedefaultendpunct}{\mcitedefaultseppunct}\relax
\EndOfBibitem
\bibitem[Goodpaster \latin{et~al.}(2014)Goodpaster, Barnes, Manby, and
  Miller~III]{goodpaster2014accurate}
Goodpaster,~J.~D.; Barnes,~T.~A.; Manby,~F.~R.; Miller~III,~T.~F. Accurate and
  systematically improvable density functional theory embedding for correlated
  wavefunctions. \emph{J. Chem. Phys.} \textbf{2014}, \emph{140}, 18A507\relax
\mciteBstWouldAddEndPuncttrue
\mciteSetBstMidEndSepPunct{\mcitedefaultmidpunct}
{\mcitedefaultendpunct}{\mcitedefaultseppunct}\relax
\EndOfBibitem
\bibitem[Manby \latin{et~al.}(2012)Manby, Stella, Goodpaster, and
  Miller~III]{manby2012simple}
Manby,~F.~R.; Stella,~M.; Goodpaster,~J.~D.; Miller~III,~T.~F. A simple, exact
  density-functional-theory embedding scheme. \emph{J. Chem. Theory Comput.}
  \textbf{2012}, \emph{8}, 2564--2568\relax
\mciteBstWouldAddEndPuncttrue
\mciteSetBstMidEndSepPunct{\mcitedefaultmidpunct}
{\mcitedefaultendpunct}{\mcitedefaultseppunct}\relax
\EndOfBibitem
\bibitem[Goodpaster \latin{et~al.}(2010)Goodpaster, Ananth, Manby, and
  Miller~III]{goodpaster2010exact}
Goodpaster,~J.~D.; Ananth,~N.; Manby,~F.~R.; Miller~III,~T.~F. Exact
  nonadditive kinetic potentials for embedded density functional theory.
  \emph{J. Chem. Phys.} \textbf{2010}, \emph{133}, 084103\relax
\mciteBstWouldAddEndPuncttrue
\mciteSetBstMidEndSepPunct{\mcitedefaultmidpunct}
{\mcitedefaultendpunct}{\mcitedefaultseppunct}\relax
\EndOfBibitem
\bibitem[Zhang \latin{et~al.}(2020)Zhang, Ren, and Caricato]{zhang2020multi}
Zhang,~K.; Ren,~S.; Caricato,~M. Multi-state QM/QM Extrapolation of UV/Vis
  Absorption Spectra with Point Charge Embedding. \emph{J. Chem. Theory
  Comput.} \textbf{2020}, \emph{16}, 4361--4372\relax
\mciteBstWouldAddEndPuncttrue
\mciteSetBstMidEndSepPunct{\mcitedefaultmidpunct}
{\mcitedefaultendpunct}{\mcitedefaultseppunct}\relax
\EndOfBibitem
\bibitem[Ramos \latin{et~al.}(2015)Ramos, Papadakis, and
  Pavanello]{ramos2015performance}
Ramos,~P.; Papadakis,~M.; Pavanello,~M. Performance of frozen density embedding
  for modeling hole transfer reactions. \emph{J. Phys. Chem. B} \textbf{2015},
  \emph{119}, 7541--7557\relax
\mciteBstWouldAddEndPuncttrue
\mciteSetBstMidEndSepPunct{\mcitedefaultmidpunct}
{\mcitedefaultendpunct}{\mcitedefaultseppunct}\relax
\EndOfBibitem
\bibitem[Pavanello and Neugebauer(2011)Pavanello, and
  Neugebauer]{pavanello2011modelling}
Pavanello,~M.; Neugebauer,~J. Modelling charge transfer reactions with the
  frozen density embedding formalism. \emph{J. Chem. Phys.} \textbf{2011},
  \emph{135}, 234103\relax
\mciteBstWouldAddEndPuncttrue
\mciteSetBstMidEndSepPunct{\mcitedefaultmidpunct}
{\mcitedefaultendpunct}{\mcitedefaultseppunct}\relax
\EndOfBibitem
\bibitem[Bennie \latin{et~al.}(2017)Bennie, Curchod, Manby, and
  Glowacki]{bennie2017pushing}
Bennie,~S.~J.; Curchod,~B.~F.; Manby,~F.~R.; Glowacki,~D.~R. Pushing the limits
  of EOM-CCSD with projector-based embedding for excitation energies. \emph{J.
  Phys. Chem. Lett.} \textbf{2017}, \emph{8}, 5559--5565\relax
\mciteBstWouldAddEndPuncttrue
\mciteSetBstMidEndSepPunct{\mcitedefaultmidpunct}
{\mcitedefaultendpunct}{\mcitedefaultseppunct}\relax
\EndOfBibitem
\bibitem[Lee \latin{et~al.}(2019)Lee, Welborn, Manby, and
  Miller~III]{lee2019projection}
Lee,~S.~J.; Welborn,~M.; Manby,~F.~R.; Miller~III,~T.~F. Projection-based
  wavefunction-in-DFT embedding. \emph{Accounts of chemical research}
  \textbf{2019}, \emph{52}, 1359--1368\relax
\mciteBstWouldAddEndPuncttrue
\mciteSetBstMidEndSepPunct{\mcitedefaultmidpunct}
{\mcitedefaultendpunct}{\mcitedefaultseppunct}\relax
\EndOfBibitem
\bibitem[Neugebauer \latin{et~al.}(2005)Neugebauer, Louwerse, Baerends, and
  Wesolowski]{neugebauer2005merits}
Neugebauer,~J.; Louwerse,~M.~J.; Baerends,~E.~J.; Wesolowski,~T.~A. The merits
  of the frozen-density embedding scheme to model solvatochromic shifts.
  \emph{J. Chem. Phys.} \textbf{2005}, \emph{122}, 094115\relax
\mciteBstWouldAddEndPuncttrue
\mciteSetBstMidEndSepPunct{\mcitedefaultmidpunct}
{\mcitedefaultendpunct}{\mcitedefaultseppunct}\relax
\EndOfBibitem
\bibitem[Wesolowski \latin{et~al.}(2015)Wesolowski, Shedge, and
  Zhou]{wesolowski2015frozen}
Wesolowski,~T.~A.; Shedge,~S.; Zhou,~X. Frozen-density embedding strategy for
  multilevel simulations of electronic structure. \emph{Chem. Rev.}
  \textbf{2015}, \emph{115}, 5891--5928\relax
\mciteBstWouldAddEndPuncttrue
\mciteSetBstMidEndSepPunct{\mcitedefaultmidpunct}
{\mcitedefaultendpunct}{\mcitedefaultseppunct}\relax
\EndOfBibitem
\bibitem[Fux \latin{et~al.}(2010)Fux, Jacob, Neugebauer, Visscher, and
  Reiher]{fux2010accurate}
Fux,~S.; Jacob,~C.~R.; Neugebauer,~J.; Visscher,~L.; Reiher,~M. Accurate
  frozen-density embedding potentials as a first step towards a subsystem
  description of covalent bonds. \emph{J. Chem. Phys.} \textbf{2010},
  \emph{132}, 164101\relax
\mciteBstWouldAddEndPuncttrue
\mciteSetBstMidEndSepPunct{\mcitedefaultmidpunct}
{\mcitedefaultendpunct}{\mcitedefaultseppunct}\relax
\EndOfBibitem
\bibitem[Jacob \latin{et~al.}(2008)Jacob, Neugebauer, and
  Visscher]{jacob2008flexible}
Jacob,~C.~R.; Neugebauer,~J.; Visscher,~L. A flexible implementation of
  frozen-density embedding for use in multilevel simulations. \emph{J. Comput.
  Chem.} \textbf{2008}, \emph{29}, 1011--1018\relax
\mciteBstWouldAddEndPuncttrue
\mciteSetBstMidEndSepPunct{\mcitedefaultmidpunct}
{\mcitedefaultendpunct}{\mcitedefaultseppunct}\relax
\EndOfBibitem
\bibitem[Jacob and Visscher(2006)Jacob, and Visscher]{jacob2006calculation}
Jacob,~C.~R.; Visscher,~L. Calculation of nuclear magnetic resonance shieldings
  using frozen-density embedding. \emph{J. Chem. Phys.} \textbf{2006},
  \emph{125}, 194104\relax
\mciteBstWouldAddEndPuncttrue
\mciteSetBstMidEndSepPunct{\mcitedefaultmidpunct}
{\mcitedefaultendpunct}{\mcitedefaultseppunct}\relax
\EndOfBibitem
\bibitem[Jacob \latin{et~al.}(2006)Jacob, Neugebauer, Jensen, and
  Visscher]{jacob2006comparison}
Jacob,~C.~R.; Neugebauer,~J.; Jensen,~L.; Visscher,~L. Comparison of
  frozen-density embedding and discrete reaction field solvent models for
  molecular properties. \emph{Phys. Chem. Chem. Phys.} \textbf{2006}, \emph{8},
  2349--2359\relax
\mciteBstWouldAddEndPuncttrue
\mciteSetBstMidEndSepPunct{\mcitedefaultmidpunct}
{\mcitedefaultendpunct}{\mcitedefaultseppunct}\relax
\EndOfBibitem
\bibitem[S{\ae}ther \latin{et~al.}(2017)S{\ae}ther, Kj{\ae}rgaard, Koch, and
  H{\o}yvik]{saether2017density}
S{\ae}ther,~S.; Kj{\ae}rgaard,~T.; Koch,~H.; H{\o}yvik,~I.-M. Density-Based
  Multilevel Hartree--Fock Model. \emph{J. Chem. Theory Comput.} \textbf{2017},
  \emph{13}, 5282--5290\relax
\mciteBstWouldAddEndPuncttrue
\mciteSetBstMidEndSepPunct{\mcitedefaultmidpunct}
{\mcitedefaultendpunct}{\mcitedefaultseppunct}\relax
\EndOfBibitem
\bibitem[H{\o}yvik(2020)]{hoyvik2020convergence}
H{\o}yvik,~I.-M. Convergence acceleration for the multilevel Hartree--Fock
  model. \emph{Mol. Phys.} \textbf{2020}, \emph{118}, 1626929\relax
\mciteBstWouldAddEndPuncttrue
\mciteSetBstMidEndSepPunct{\mcitedefaultmidpunct}
{\mcitedefaultendpunct}{\mcitedefaultseppunct}\relax
\EndOfBibitem
\bibitem[Aquilante \latin{et~al.}(2011)Aquilante, Boman, Bostr{\"o}m, Koch,
  Lindh, de~Mer{\'a}s, and Pedersen]{aquilante2011cholesky}
Aquilante,~F.; Boman,~L.; Bostr{\"o}m,~J.; Koch,~H.; Lindh,~R.;
  de~Mer{\'a}s,~A.~S.; Pedersen,~T.~B. \emph{Linear-Scaling Techniques in
  Computational Chemistry and Physics}; Springer, 2011; pp 301--343\relax
\mciteBstWouldAddEndPuncttrue
\mciteSetBstMidEndSepPunct{\mcitedefaultmidpunct}
{\mcitedefaultendpunct}{\mcitedefaultseppunct}\relax
\EndOfBibitem
\bibitem[S{\'a}nchez~de Mer{\'a}s \latin{et~al.}(2010)S{\'a}nchez~de Mer{\'a}s,
  Koch, Cuesta, and Boman]{sanchez2010cholesky}
S{\'a}nchez~de Mer{\'a}s,~A.~M.; Koch,~H.; Cuesta,~I.~G.; Boman,~L. Cholesky
  decomposition-based definition of atomic subsystems in electronic structure
  calculations. \emph{J. Chem. Phys.} \textbf{2010}, \emph{132}, 204105\relax
\mciteBstWouldAddEndPuncttrue
\mciteSetBstMidEndSepPunct{\mcitedefaultmidpunct}
{\mcitedefaultendpunct}{\mcitedefaultseppunct}\relax
\EndOfBibitem
\bibitem[Koch \latin{et~al.}(2003)Koch, S{\'a}nchez~de Mer{\'a}s, and
  Pedersen]{koch2003reduced}
Koch,~H.; S{\'a}nchez~de Mer{\'a}s,~A.; Pedersen,~T.~B. Reduced scaling in
  electronic structure calculations using Cholesky decompositions. \emph{J.
  Chem. Phys.} \textbf{2003}, \emph{118}, 9481--9484\relax
\mciteBstWouldAddEndPuncttrue
\mciteSetBstMidEndSepPunct{\mcitedefaultmidpunct}
{\mcitedefaultendpunct}{\mcitedefaultseppunct}\relax
\EndOfBibitem
\bibitem[Folkestad \latin{et~al.}(2020)Folkestad, Kjønstad, Myhre, Andersen,
  Balbi, Coriani, Giovannini, Goletto, Haugland, Hutcheson, Høyvik, Moitra,
  Paul, Scavino, Skeidsvoll, Åsmund H.~Tveten, and Koch]{eT_arxiv}
Folkestad,~S.~D.; Kjønstad,~E.~F.; Myhre,~R.~H.; Andersen,~J.~H.; Balbi,~A.;
  Coriani,~S.; Giovannini,~T.; Goletto,~L.; Haugland,~T.~S.; Hutcheson,~A.;
  Høyvik,~I.-M.; Moitra,~T.; Paul,~A.~C.; Scavino,~M.; Skeidsvoll,~A.~S.;
  Åsmund H.~Tveten,; Koch,~H. eT 1.0: an open source electronic structure
  program with emphasis on coupled cluster and multilevel methods. \emph{arXiv}
  \textbf{2020}, 2002.05631\relax
\mciteBstWouldAddEndPuncttrue
\mciteSetBstMidEndSepPunct{\mcitedefaultmidpunct}
{\mcitedefaultendpunct}{\mcitedefaultseppunct}\relax
\EndOfBibitem
\bibitem[Lebedev and Laikov(1999)Lebedev, and Laikov]{lebedev1999quadrature}
Lebedev,~V.~I.; Laikov,~D. \emph{Doklady Mathematics}; 1999; Vol.~59; pp
  477--481\relax
\mciteBstWouldAddEndPuncttrue
\mciteSetBstMidEndSepPunct{\mcitedefaultmidpunct}
{\mcitedefaultendpunct}{\mcitedefaultseppunct}\relax
\EndOfBibitem
\bibitem[Marques \latin{et~al.}(2012)Marques, Oliveira, and
  Burnus]{marques2012libxc}
Marques,~M.~A.; Oliveira,~M.~J.; Burnus,~T. Libxc: A library of exchange and
  correlation functionals for density functional theory. \emph{Comput. Phys.
  Commun.} \textbf{2012}, \emph{183}, 2272--2281\relax
\mciteBstWouldAddEndPuncttrue
\mciteSetBstMidEndSepPunct{\mcitedefaultmidpunct}
{\mcitedefaultendpunct}{\mcitedefaultseppunct}\relax
\EndOfBibitem
\bibitem[He and Merz~Jr(2010)He, and Merz~Jr]{he2010divide}
He,~X.; Merz~Jr,~K.~M. Divide and conquer Hartree- Fock calculations on
  proteins. \emph{J. Chem. Theory Comput.} \textbf{2010}, \emph{6},
  405--411\relax
\mciteBstWouldAddEndPuncttrue
\mciteSetBstMidEndSepPunct{\mcitedefaultmidpunct}
{\mcitedefaultendpunct}{\mcitedefaultseppunct}\relax
\EndOfBibitem
\bibitem[Kohn and Sham(1965)Kohn, and Sham]{kohn1965self}
Kohn,~W.; Sham,~L.~J. Self-consistent equations including exchange and
  correlation effects. \emph{Phys. Rev.} \textbf{1965}, \emph{140}, A1133\relax
\mciteBstWouldAddEndPuncttrue
\mciteSetBstMidEndSepPunct{\mcitedefaultmidpunct}
{\mcitedefaultendpunct}{\mcitedefaultseppunct}\relax
\EndOfBibitem
\bibitem[Perdew \latin{et~al.}(1997)Perdew, Burke, and
  Ernzerhof]{perdew1997generalized}
Perdew,~J.~P.; Burke,~K.; Ernzerhof,~M. Generalized gradient approximation made
  simple. \emph{Phys. Rev. Lett.} \textbf{1997}, \emph{77}, 3865\relax
\mciteBstWouldAddEndPuncttrue
\mciteSetBstMidEndSepPunct{\mcitedefaultmidpunct}
{\mcitedefaultendpunct}{\mcitedefaultseppunct}\relax
\EndOfBibitem
\bibitem[Becke(1993)]{becke1993density}
Becke,~A.~D. Density-functional thermochemistry. III. The role of exact
  exchange. \emph{J. Chem. Phys.} \textbf{1993}, \emph{98}, 5648--5652\relax
\mciteBstWouldAddEndPuncttrue
\mciteSetBstMidEndSepPunct{\mcitedefaultmidpunct}
{\mcitedefaultendpunct}{\mcitedefaultseppunct}\relax
\EndOfBibitem
\bibitem[Giovannini \latin{et~al.}(2020)Giovannini, Egidi, and
  Cappelli]{giovannini2020csr}
Giovannini,~T.; Egidi,~F.; Cappelli,~C. Molecular Spectroscopy of Aqueous
  Solutions: A The- oretical Perspective. \emph{Chem. Soc. Rev.} \textbf{2020},
  DOI: 10.1039/c9cs00464e\relax
\mciteBstWouldAddEndPuncttrue
\mciteSetBstMidEndSepPunct{\mcitedefaultmidpunct}
{\mcitedefaultendpunct}{\mcitedefaultseppunct}\relax
\EndOfBibitem
\bibitem[Giovannini \latin{et~al.}(2019)Giovannini, Riso, Ambrosetti, Puglisi,
  and Cappelli]{giovannini2019fqfmulinear}
Giovannini,~T.; Riso,~R.~R.; Ambrosetti,~M.; Puglisi,~A.; Cappelli,~C.
  Electronic transitions for a fully polarizable qm/mm approach based on
  fluctuating charges and fluctuating dipoles: linear and corrected linear
  response regimes. \emph{J. Chem. Phys.} \textbf{2019}, \emph{151},
  174104\relax
\mciteBstWouldAddEndPuncttrue
\mciteSetBstMidEndSepPunct{\mcitedefaultmidpunct}
{\mcitedefaultendpunct}{\mcitedefaultseppunct}\relax
\EndOfBibitem
\bibitem[Giovannini \latin{et~al.}(2018)Giovannini, Del~Frate, Lafiosca, and
  Cappelli]{giovannini2018effective}
Giovannini,~T.; Del~Frate,~G.; Lafiosca,~P.; Cappelli,~C. Effective
  computational route towards vibrational optical activity spectra of chiral
  molecules in aqueous solution. \emph{Phys. Chem. Chem. Phys.} \textbf{2018},
  \emph{20}, 9181--9197\relax
\mciteBstWouldAddEndPuncttrue
\mciteSetBstMidEndSepPunct{\mcitedefaultmidpunct}
{\mcitedefaultendpunct}{\mcitedefaultseppunct}\relax
\EndOfBibitem
\bibitem[Lipparini \latin{et~al.}(2013)Lipparini, Cappelli, and
  Barone]{lipparini2013gauge}
Lipparini,~F.; Cappelli,~C.; Barone,~V. A gauge invariant multiscale approach
  to magnetic spectroscopies in condensed phase: General three-layer model,
  computational implementation and pilot applications. \emph{J. Chem. Phys.}
  \textbf{2013}, \emph{138}, 234108\relax
\mciteBstWouldAddEndPuncttrue
\mciteSetBstMidEndSepPunct{\mcitedefaultmidpunct}
{\mcitedefaultendpunct}{\mcitedefaultseppunct}\relax
\EndOfBibitem
\bibitem[Giovannini \latin{et~al.}(2019)Giovannini, Grazioli, Ambrosetti, and
  Cappelli]{giovannini2019fqfmuir}
Giovannini,~T.; Grazioli,~L.; Ambrosetti,~M.; Cappelli,~C. Calculation of ir
  spectra with a fully polarizable qm/mm approach based on fluctuating charges
  and fluctuating dipoles. \emph{J. Chem. Theory Comput.} \textbf{2019},
  \emph{15}, 5495--5507\relax
\mciteBstWouldAddEndPuncttrue
\mciteSetBstMidEndSepPunct{\mcitedefaultmidpunct}
{\mcitedefaultendpunct}{\mcitedefaultseppunct}\relax
\EndOfBibitem
\bibitem[Giovannini \latin{et~al.}(2017)Giovannini, Olsz{\`o}wka, Egidi,
  Cheeseman, Scalmani, and Cappelli]{giovannini2017polarizable}
Giovannini,~T.; Olsz{\`o}wka,~M.; Egidi,~F.; Cheeseman,~J.~R.; Scalmani,~G.;
  Cappelli,~C. Polarizable Embedding Approach for the Analytical Calculation of
  Raman and Raman Optical Activity Spectra of Solvated Systems. \emph{J. Chem.
  Theory Comput.} \textbf{2017}, \emph{13}, 4421--4435\relax
\mciteBstWouldAddEndPuncttrue
\mciteSetBstMidEndSepPunct{\mcitedefaultmidpunct}
{\mcitedefaultendpunct}{\mcitedefaultseppunct}\relax
\EndOfBibitem
\bibitem[Losada \latin{et~al.}(2008)Losada, Nguyen, and
  Xu]{losada2008solvation}
Losada,~M.; Nguyen,~P.; Xu,~Y. Solvation of propylene oxide in water:
  Vibrational circular dichroism, optical rotation, and computer simulation
  studies. \emph{J. Phys. Chem A} \textbf{2008}, \emph{112}, 5621--5627\relax
\mciteBstWouldAddEndPuncttrue
\mciteSetBstMidEndSepPunct{\mcitedefaultmidpunct}
{\mcitedefaultendpunct}{\mcitedefaultseppunct}\relax
\EndOfBibitem
\bibitem[Yang and Xu(2009)Yang, and Xu]{yang2009probing}
Yang,~G.; Xu,~Y. Probing chiral solute-water hydrogen bonding networks by
  chirality transfer effects: a vibrational circular dichroism study of
  glycidol in water. \emph{J. Chem. Phys.} \textbf{2009}, \emph{130},
  164506--164506\relax
\mciteBstWouldAddEndPuncttrue
\mciteSetBstMidEndSepPunct{\mcitedefaultmidpunct}
{\mcitedefaultendpunct}{\mcitedefaultseppunct}\relax
\EndOfBibitem
\bibitem[Merten \latin{et~al.}(2013)Merten, Bloino, Barone, and
  Xu]{merten2013anharmonicity}
Merten,~C.; Bloino,~J.; Barone,~V.; Xu,~Y. Anharmonicity Effects in the
  Vibrational {CD} Spectra of Propylene Oxide. \emph{J. Phys. Chem.. Lett.}
  \textbf{2013}, \emph{4}, 3424--3428\relax
\mciteBstWouldAddEndPuncttrue
\mciteSetBstMidEndSepPunct{\mcitedefaultmidpunct}
{\mcitedefaultendpunct}{\mcitedefaultseppunct}\relax
\EndOfBibitem
\bibitem[Su and Xu(2007)Su, and Xu]{su2007hydration}
Su,~Z.; Xu,~Y. Hydration of a Chiral Molecule: The Propylene
  Oxide$\cdots$(Water)$_2$ Cluster in the Gas Phase. \emph{Anew. Chem. Int.
  Edit.} \textbf{2007}, \emph{119}, 6275--6278\relax
\mciteBstWouldAddEndPuncttrue
\mciteSetBstMidEndSepPunct{\mcitedefaultmidpunct}
{\mcitedefaultendpunct}{\mcitedefaultseppunct}\relax
\EndOfBibitem
\bibitem[Yu \latin{et~al.}(2009)Yu, Xu, Wei, Wang, He, Xia, Zhang, and
  Liu]{yu2009new}
Yu,~Z.; Xu,~L.; Wei,~Y.; Wang,~Y.; He,~Y.; Xia,~Q.; Zhang,~X.; Liu,~Z. A new
  route for the synthesis of propylene oxide from bio-glycerol derivated
  propylene glycol. \emph{Chem. Comm.} \textbf{2009}, 3934--3936\relax
\mciteBstWouldAddEndPuncttrue
\mciteSetBstMidEndSepPunct{\mcitedefaultmidpunct}
{\mcitedefaultendpunct}{\mcitedefaultseppunct}\relax
\EndOfBibitem
\bibitem[Su \latin{et~al.}(2006)Su, Borho, and Xu]{su2006chiral}
Su,~Z.; Borho,~N.; Xu,~Y. Chiral self-recognition: Direct spectroscopic
  detection of the homochiral and heterochiral dimers of propylene oxide in the
  gas phase. \emph{J. Am. Chem. Soc.} \textbf{2006}, \emph{128},
  17126--17131\relax
\mciteBstWouldAddEndPuncttrue
\mciteSetBstMidEndSepPunct{\mcitedefaultmidpunct}
{\mcitedefaultendpunct}{\mcitedefaultseppunct}\relax
\EndOfBibitem
\bibitem[Su \latin{et~al.}(2006)Su, Wen, and Xu]{su2006conformational}
Su,~Z.; Wen,~Q.; Xu,~Y. Conformational Stability of the Propylene Oxide- Water
  Adduct: Direct Spectroscopic Detection of O- H$\cdots$ O Hydrogen Bonded
  Conformers. \emph{J. Am. Chem. Soc.} \textbf{2006}, \emph{128},
  6755--6760\relax
\mciteBstWouldAddEndPuncttrue
\mciteSetBstMidEndSepPunct{\mcitedefaultmidpunct}
{\mcitedefaultendpunct}{\mcitedefaultseppunct}\relax
\EndOfBibitem
\bibitem[Perera \latin{et~al.}(2016)Perera, Thomas, Poopari, and
  Xu]{perera2016clusters}
Perera,~A.~S.; Thomas,~J.; Poopari,~M.~R.; Xu,~Y. The clusters-in-a-liquid
  approach for solvation: new insights from the conformer specific gas phase
  spectroscopy and vibrational optical activity spectroscopy. \emph{Front.
  Chem.} \textbf{2016}, \emph{4}, 9\relax
\mciteBstWouldAddEndPuncttrue
\mciteSetBstMidEndSepPunct{\mcitedefaultmidpunct}
{\mcitedefaultendpunct}{\mcitedefaultseppunct}\relax
\EndOfBibitem
\bibitem[Lipparini \latin{et~al.}(2013)Lipparini, Egidi, Cappelli, and
  Barone]{lipparini2013optical}
Lipparini,~F.; Egidi,~F.; Cappelli,~C.; Barone,~V. The optical rotation of
  methyloxirane in aqueous solution: a never ending story? \emph{J. Chem.
  Theory Comput.} \textbf{2013}, \emph{9}, 1880--1884\relax
\mciteBstWouldAddEndPuncttrue
\mciteSetBstMidEndSepPunct{\mcitedefaultmidpunct}
{\mcitedefaultendpunct}{\mcitedefaultseppunct}\relax
\EndOfBibitem
\bibitem[Giovannini \latin{et~al.}(2016)Giovannini, Olszowka, and
  Cappelli]{giovannini2016effective}
Giovannini,~T.; Olszowka,~M.; Cappelli,~C. Effective Fully Polarizable QM/MM
  Approach To Model Vibrational Circular Dichroism Spectra of Systems in
  Aqueous Solution. \emph{J. Chem. Theory Comput.} \textbf{2016}, \emph{12},
  5483--5492\relax
\mciteBstWouldAddEndPuncttrue
\mciteSetBstMidEndSepPunct{\mcitedefaultmidpunct}
{\mcitedefaultendpunct}{\mcitedefaultseppunct}\relax
\EndOfBibitem
\end{mcitethebibliography}
}

\end{document}